# Discovery and Synthesis of a Family of Boride Altermagnets[§]

Zhen Zhang[1,#], Eranga H. Gamage[2,3,#], Genevieve Amobi[2], Subhadip Pradhan[4], Andrey Kutepov[4], Kirill D. Belashchenko[4], Yang Sun[5], Kirill Kovnir[2,3,*], and Vladimir Antropov[1,3,*]

[1]Department of Physics and Astronomy, Iowa State University, Ames, IA 50011, USA
[2]Department of Chemistry, Iowa State University, Ames, IA 50011, USA
[3]Ames National Laboratory, U.S. Department of Energy, Ames, IA 50011, USA
[4]Department of Physics and Astronomy and Nebraska Center for Materials and Nanoscience, University of Nebraska-Lincoln, Lincoln, NE 68588, USA
[5]Department of Physics, Xiamen University, Xiamen 361005, China
[#]These authors contributed equally.
[*]Corresponding authors: Kirill Kovnir kovnir@iastate.edu, Vladimir Antropov antropov@iastate.edu

**Abstract**

Borides are a rich material family. To push the boundaries of borides' properties and applications into broader fields, we have conducted systematic theoretical and experimental searches for synthesizable phases in ternary borides $TM_2B_2$ ($T$ = 3$d$, $M$ = 4$d$/5$d$ transition metals). We find that $TM_2B_2$ in the $FeMo_2B_2$-type and $CoW_2B_2$-type structures form a large family of stable/metastable materials of 120 members. Among them, we identify 40 materials with stable magnetic solutions. Further, we discover 11 altermagnets in the $FeMo_2B_2$-type structure. So far, boride altermagnets are rare. In these altermagnets, $T$ = Fe or Mn atoms are arranged in parallel $T$-chains with strong ferromagnetic intrachain couplings and antiferromagnetic interchain couplings. They simultaneously exhibit electronic band spin splitting, typical of ferromagnetism, and zero net magnetization, typical of antiferromagnetism. They also exhibit magnonic band chiral splitting. Both effects originate from the unique altermagnetic symmetries crucially constrained by the nonmagnetic atoms in the structure. Transport properties of relevance to spintronic applications, including the strain-induced spin-splitter effect and anomalous Hall effect, are predicted. An iodine-assisted synthesis method for $TM_2B_2$ is developed, using which 7 of the predicted low-energy phases are experimentally synthesized and characterized, including 4 altermagnets. This work expands the realm of borides by offering new opportunities for studying altermagnetism and altermagnons in borides. It also provides valuable insights into the discovery and design of altermagnets. By demonstrating that altermagnets can exist as families sharing a common motif, this work paves a feasible route for discovering altermagnets by elemental substitutions and high-throughput computations.

**INTRODUCTION**

Borides are renowned for their diverse structures and exceptional properties, making them highly versatile across multiple applications. Borides are also recognized for the key role played by boron as a structure stabilizer in some permanent magnets, e.g., the top-performing $Nd_2Fe_{14}B$[1], though boron itself is nonmagnetic. The capability of boron to form binary, ternary, or multinary compounds with most metals, coupled with their wide variety of atomic arrangements[2], underscores borides' extensive potential in conventional magnetism[3–6], superhardness[7–9], superconductivity[10–12], electrochemistry[13–16], laminate and two-dimensional (2D) materials[17–23].

The abundance and the structural and compositional diversity of borides create difficulties for efficient experimental exploration of novel borides, which limit the development of borides' potentially broader applications. Theoretical computations, on the other hand, can make up for the shortcomings of experiments, accelerating the discovery of new synthesizable and functional materials. Recently, by combining systematic screening and theoretical analysis with experimental efforts, novel quantum magnetism[24,25] has been unveiled in the borides family. A wider range of previously unknown properties and applications is anticipated to be found in borides.

Motivated by discovering novel magnetic borides, we performed comprehensive theoretical and experimental studies for a ternary transition metal boride family with 1-2-2 chemical composition. This family of materials has various structures and elemental compositions. Several experimental observations for them were briefly mentioned more than half a century ago[26–28]. However, so far, systematic theoretical and experimental evaluations of their stability and magnetism are scarce. Here, we identified a substantial number of stable and metastable compounds in this family, including altermagnets, conventional magnets, and nonmagnets. Altermagnetism is a recently discovered third fundamental type of collinear magnetism besides ferromagnetism and antiferromagnetism[29–32]. It simultaneously exhibits electronic band spin splitting, typical of ferromagnetism, and zero net magnetization, typical of antiferromagnetism. So far, altermagnets have been found mainly in compounds between transition metals and group 4A–7A elements[30] but not in borides.

In this work, we used first-principles calculations to thoroughly study the $TM_2B_2$ compounds, where $T$ is magnetic $3d$ transition metals, V–Ni, and $M$ is nonmagnetic $4d$ and $5d$ transition metals, Y–Cd and La–Hg ($4f$ elements Ce–Lu were omitted). We found 120 thermodynamically favorable phases, including 11 altermagnets and 29 conventional magnets. To verify the predicted thermodynamic stabilities, structural features, and magnetic behaviors, we developed a new solid-state synthesis method capable of producing high-purity samples below 1100 °C within 24 hours. Using the new synthesis method, we experimentally synthesized 7 thermodynamically favorable ternary compounds, including 4 altermagnets.

**RESULTS AND DISCUSSION**

**Crystal Structures.** We were most interested in the $TM_2B_2$ phases with the $FeMo_2B_2$-type (tetragonal, space group $P4/mbm$)[26] structure. The $FeMo_2B_2$-type tetragonal structure is displayed in Figures 1a and 1b. It crystallizes in the $P4/mbm$ space group with a 10-atom unit cell. The crystal structure features a 2D layered arrangement of $T$-B atoms with $M$ atoms between the layers (Figure 1a). The layers are composed of $T_2B_3$ pentagons with B-B bonds in the $ab$ plane. $M$ atoms are located at the center of the pentagonal prisms (Figure 1b). The layers are stacked on top of each other along the $c$-axis. The nearest neighbors between $T$ atoms are along the $c$-axis, forming a $T$-chain with equally spaced $T$ atoms.

In this tetragonal structure, the $T$ sites can also be occupied by $4d$ and $5d$ transition metals. $T = 4d$ and $5d$ are expected to switch from $T$-chain to $T$-dumbbells due to their larger radius than $3d$, as seen in RuNb$_2$B$_2$ and OsNb$_2$B$_2$[33–35]. This leads to a different OsNb$_2$B$_2$-type (tetragonal, space group $P4/mnc$) structure. Since this study is focused on magnetic $T = 3d$ elements, we do not consider the OsNb$_2$B$_2$-type structure.

Besides the FeMo$_2$B$_2$-type phase, there is a competing $TM_2B_2$ phase with the CoW$_2$B$_2$-type (orthorhombic, space group $Immm$) structure[36]. To evaluate the global stability of the $TM_2B_2$ phases, the CoW$_2$B$_2$-type structure must be taken into account, both theoretically and experimentally, in addition to the FeMo$_2$B$_2$-type structure. The CoW$_2$B$_2$-type orthorhombic structure is displayed in Figures 1c and 1d. It crystallizes in the $Immm$ space group, which also has 10 atoms per unit cell. The 2D layers extending parallel to the $bc$ plane consist of $T_2B_4$ hexagons with B-B bonds and $T_2B_2$ rhombi (Figure 1c). The $M$ atoms are located between the $T$-B layers inside ten-vertex polyhedra, each formed by a $T_2B_4$ hexagon and a $T_2B_2$ rhombus lying in parallel planes (Figure 1d). The layers are stacked along the $a$-axis in an $ABAB$ fashion.

**Phase Stability.** For each $T$-$M$-B ternary system, the phase stabilities of both structures were evaluated by calculating the ternary convex hull. The known reference phases on the convex hull were obtained from the Materials Project[37] database. The formation energy ($E_{form}$) of each phase was computed using spin-polarized density-functional theory (DFT). All phases, including the reference ones and the two constructed ones, were fully relaxed, and the total energies were fully recalculated using the same DFT settings in this work. Details of computational methods can be found in Text S1 of the Supporting Information. The formation energy relative to the convex hull ($E_d$) was then evaluated by the $E_{form}$ differences with respect to the three reference phases that form the Gibbs triangle on the convex hull. When constructing the convex hull, if a $TM_2B_2$ phase has $E_d > 0$, then it is above the known convex hull. In this case, the convex hull does not need to be updated. If a $TM_2B_2$ phase has $E_d < 0$, then it is below the known convex hull. In this case, the convex hull is updated to include the $TM_2B_2$'s $E_{form}$, and the $TM_2B_2$ is denoted as $E_d = 0$, i.e., a stable phase and a part of the new convex hull.

The calculated $E_d$ of all $TM_2B_2$ are shown in Figures 1e and 1f. The tetragonal and the orthorhombic structures are represented by blue and red bars, respectively. The $E_d$ of the two structures for the same composition do not differ significantly. They show systematic peaks and dips in the trend of $E_d$ as the electron count varies along the periodic table. The variation of the $3d$ $T$ atom does not change $E_d$ significantly. Instead, the $4d$ or $5d$ $M$ atom largely determines where $E_d$ peaks and dips. When $M$ is Zr, Nb, Mo, Tc, or their heavier homologs Hf, Ta, W, and Re, the $TM_2B_2$ phases are usually stable or close to stable.

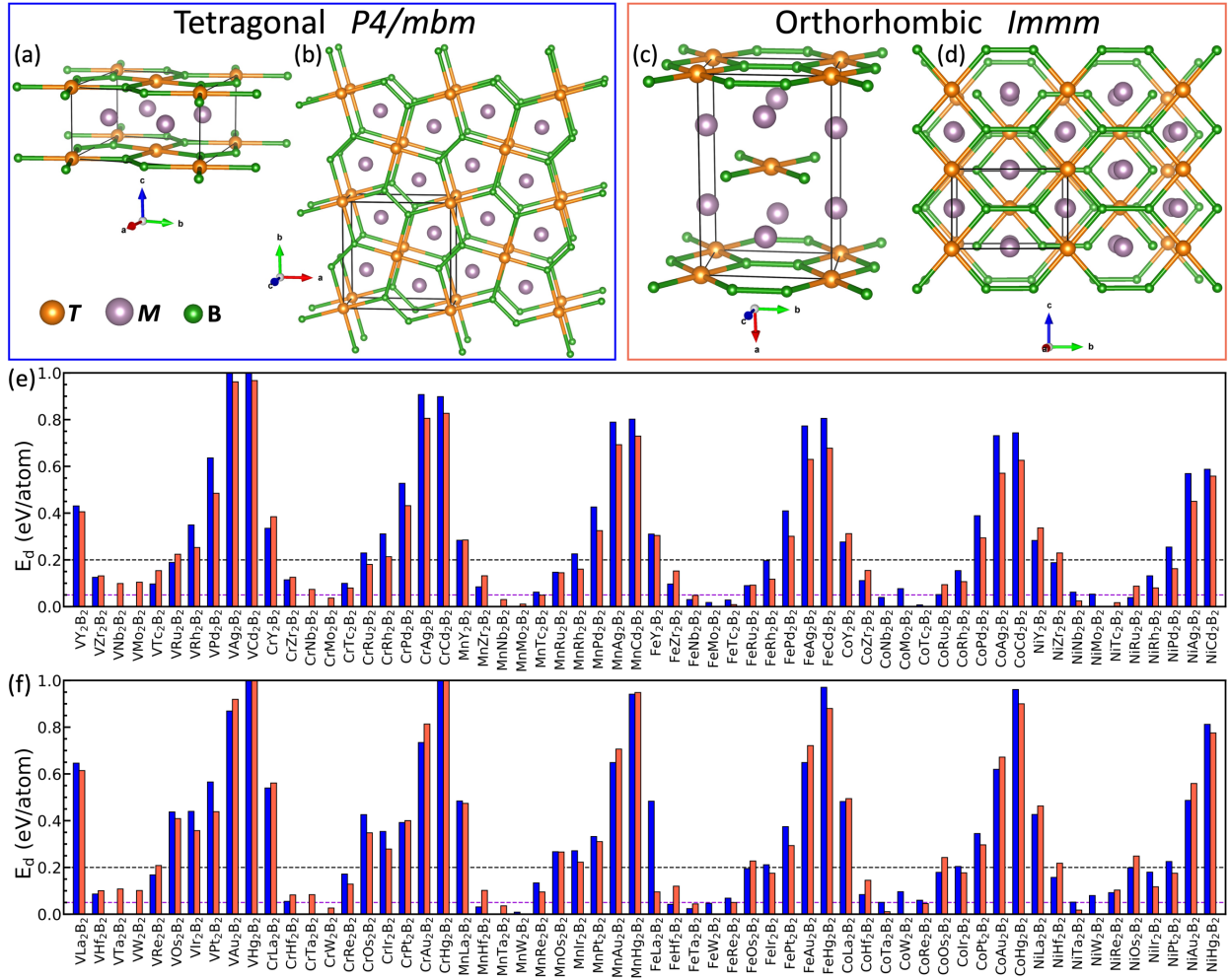

**Figure 1.** Crystal structure and hull distance of $TM_2B_2$ phases. (a) General view and (b) view along [001] direction of the FeMo$_2$B$_2$-type tetragonal *P4/mbm* unit cell. *T* (orange): 3*d* transition metals; *M* (purple): 4*d* or 5*d* transition metals; B (green): boron. (c) General view and (d) view along [100] direction of the CoW$_2$B$_2$-type orthorhombic *Immm* unit cell. $TM_2B_2$ phases' hull distance for *M* = (e) 4*d* and (f) 5*d* transition metals. The FeMo$_2$B$_2$-type and the CoW$_2$B$_2$-type phases are indicated by the blue and red bars, respectively. The criteria of $E_d \leq 0.2$ eV/atom and $E_d \leq 0.05$ eV/atom are indicated by the black and purple horizontal dashed lines, respectively.

Then, we used 0.2 eV/atom[38] as a criterion for choosing metastable compounds that are possibly stabilized by thermodynamics. In fact, some novel ternary borides with a prediction of $E_d$ = 0.21 eV/atom have been synthesized in experiments[19,20]. This way, we obtained 60 stable/metastable and 60 unstable compounds in the FeMo$_2$B$_2$-type tetragonal structure. We also obtained 60 stable/metastable and 60 unstable compounds in the CoW$_2$B$_2$-type orthorhombic structure. Furthermore, we used 0.05 eV/atom[39–41] as a stricter criterion to select compounds with significant synthesis potential in experiments. This way, 23 and 24 of the most promising compounds were obtained for the tetragonal and orthorhombic structures, respectively.

**Electronic Structures of the Nonmagnetic States.** The nonmagnetic density of states (DOS) of some of the lowest-energy tetragonal phases ($M$ = Nb, Mo, Ta, W) are plotted in Figure 2. Our discussion of the electronic and magnetic properties will focus on the magnetically interesting tetragonal phase. All these compounds have significant total DOS at the Fermi level $N(E_f)$, which is mainly contributed by the $T$ and $M$ atoms, while the contribution from B is negligible. In the energy range shown, the orbitals of $T$ and $M$ atoms are clearly hybridized[34,35,42,43], which results in coincident peaks in their partial DOS.

A prominent feature in the nonmagnetic partial DOS is the presence of the $T$ atom's major sharp peaks, which are caused by flat bands in their electronic band structure (Figure S1 of the Supporting Information). In general, enhanced $N(E_f)$ near the Fermi level can lead to instabilities such as magnetism, charge density waves, and phase transitions in materials; it can also promote superconductivity.

Here, for V, Cr, and Ni compounds, the Fermi level is located away from the major sharp peaks. Therefore, these compounds have $T$'s partial $N(E_f) < 1$ eV$^{-1}$ f.u.$^{-1}$ spin$^{-1}$, which does not satisfy the Stoner criterion and does not generate a magnetic instability toward ferromagnetism. For $NiNb_2B_2$ and $NiTa_2B_2$, the Fermi level is right above a major peak. Applying some hole doping may move the Fermi level toward the peak.

For Mn, Fe, and Co compounds, the Fermi level is located on or in close proximity to major sharp peaks as a result of electronic band filling, which leads to $T$'s partial $N(E_f) > 1$ eV$^{-1}$ f.u.$^{-1}$ spin$^{-1}$ for most of these compounds. The high $N(E_f)$ can give rise to various instabilities, including a magnetic transition toward ferromagnetism. Even in the case of $FeNb_2B_2$ and $FeTa_2B_2$, where two major sharp peaks straddle the Fermi level and the partial $N(E_f) < 1$ eV$^{-1}$ f.u.$^{-1}$ spin$^{-1}$, antiferromagnetism needs to be considered. For $CoMo_2B_2$ and $CoW_2B_2$, the Fermi level is located on a minor peak next to the two major peaks, and strong magnetic fluctuations may be expected. Although magnetism is generally expected in the Mn, Fe, and Co compounds here, whether stable magnetic solutions can exist needs to be checked by self-consistent magnetic calculations. We found that $FeNb_2B_2$ and $FeTa_2B_2$ are magnetic, while $CoMo_2B_2$ and $CoW_2B_2$ are not. Compounds found to have stable magnetic solutions are indicated by check marks in Figure 2. The results are generally in line with our analysis of the nonmagnetic electronic structures.

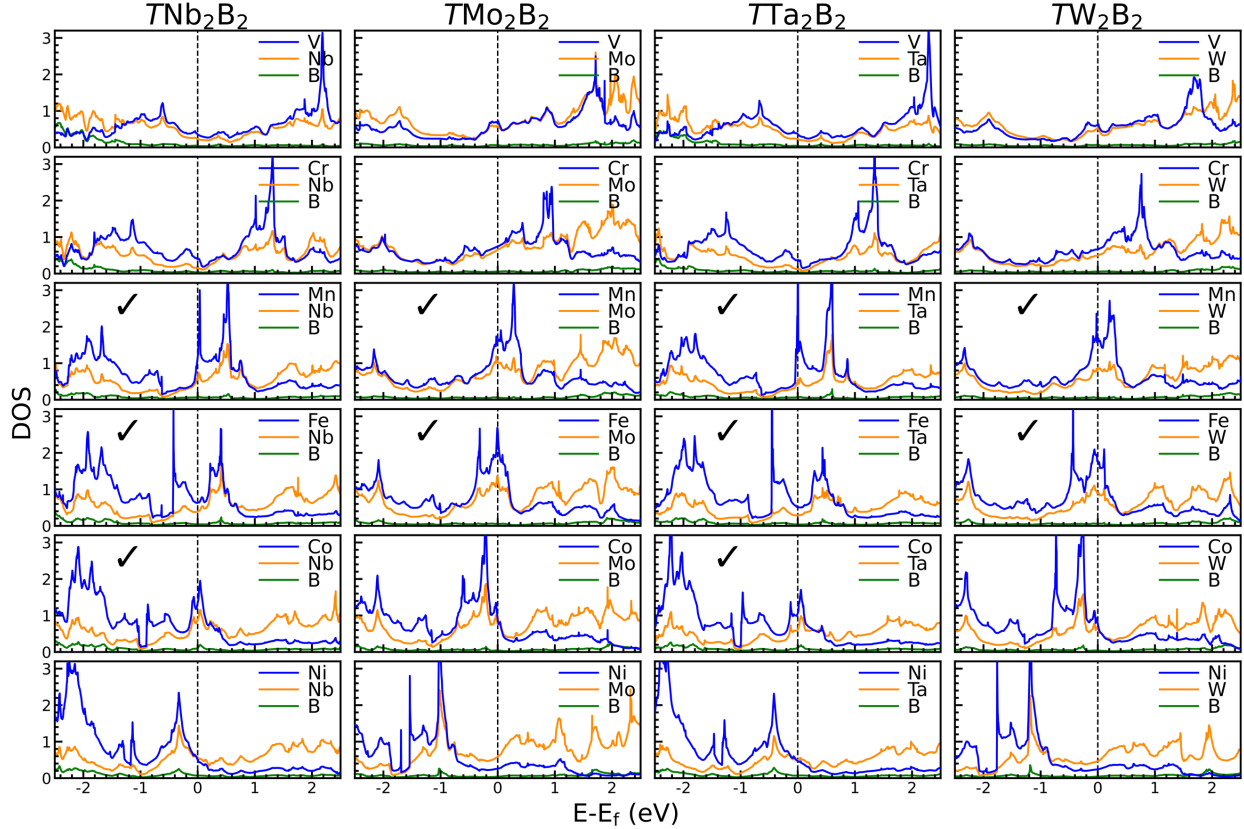

**Figure 2.** Nonmagnetic electronic density of states (eV$^{-1}$ f.u.$^{-1}$ spin$^{-1}$) of the tetragonal $TM_2B_2$ ($T$ = V, Cr, Mn, Fe, Co, Ni; $M$ = Nb, Mo, Ta, W). Blue, orange, and green curves represent the partial DOS (PDOS) of $T$, $M$, and B atoms, respectively. Check marks indicate systems with stable magnetic solutions.

**Magnetic States.** Nonmagnetic electronic structures demonstrate a general picture of possible ferromagnetic (FM) instabilities in numerous phases. To avoid leaving out antiferromagnetic (AFM) solutions, only self-consistent calculations of the magnetic ground state can provide a reliable answer. We considered all possible magnetic orderings with FM or AFM alignment of the local moments between the first, second, and third nearest $T$-$T$ neighbors. (The first three nearest neighbors are indicated in Figure 3b.) The realization of these magnetic orderings requires at least a 20-atom $1 \times 1 \times 2$ supercell, where the shortest $c$ lattice vector is doubled. Four inequivalent magnetic orderings are displayed in Figure 3b. We employed a three-letter notation using "F" for FM and "A" for AFM to represent these orderings, where each letter specifies the alignment of neighbors 1, 2, and 3 in sequential order. They were included in the spin-polarized and nonrelativistic calculations for all the $TM_2B_2$ phases.

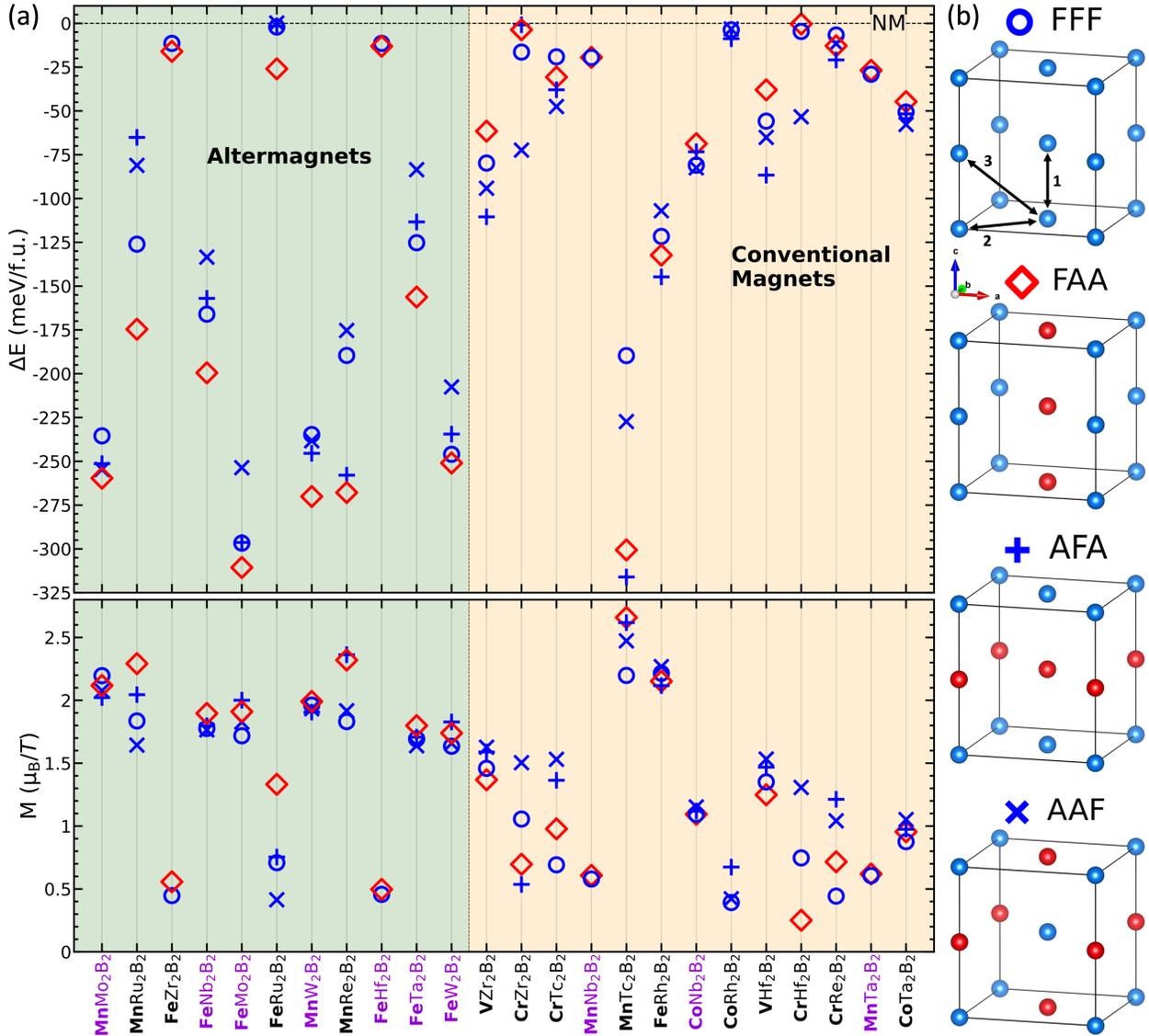

**Figure 3.** Magnetic solutions for the tetragonal $TM_2B_2$ phases within 0.2 eV/atom above the convex hull. (a) The relative energy difference of the magnetic solutions with respect to the nonmagnetic one (top panel) and the magnetic moment on the $T$ atom (bottom panel). Purple compositional x-labels indicate compounds within 0.05 eV/atom above the convex hull. Black compositional x-labels indicate compounds within 0.2 eV/atom above the convex hull. (b) Different magnetic configurations within the 20-atom $1 \times 1 \times 2$ supercell and the associated symbols and labels. Blue and red spheres indicate $T$ atoms with opposite spins. $M$ and B atoms are not shown.

Figure 3a shows all the stable magnetic solutions for the stable/metastable phases within 0.2 eV/atom above the convex hull. We found a total of 24 magnetic compounds, the remaining 36 compounds being nonmagnetic. The upper panel of Figure 3a shows the relative magnetic energy of different magnetic configurations to the nonmagnetic state, and the lower panel shows the magnetic moment on the $T$ atom. The absence of certain magnetic configurations indicates the absence of DFT self-consistent solutions with finite moments. The configuration with the lowest energy in the upper panel of Figure 3a indicates the magnetic ground state for each compound. The thermodynamically most promising phases within 0.05 eV/atom above the convex hull are highlighted by the purple compositional x-labels. There are 10 magnetic compounds with great synthesis potential.

Similarly, we considered all possible FM and AFM orderings within a 20-atom $1 \times 1 \times 2$ supercell (the shortest $c$ lattice vector is doubled) for the orthorhombic structure, and three inequivalent magnetic configurations were obtained. Screening all the orthorhombic phases by self-consistent DFT calculations yielded 16 magnetic and 44 nonmagnetic compounds within 0.2 eV/atom above the convex hull. The magnetic solutions are exhibited in Figure S2. A comprehensive classification of magnetic or nonmagnetic phases and their energetics for the two structures is summarized in Table 1.

Focusing on the tetragonal phase, only the FFF ordering (blue circles in Figure 3) is FM with non-zero net magnetization. The other three orderings have zero net magnetization. $MnNb_2B_2$, $CoNb_2B_2$, $MnTa_2B_2$, and $FeHf_2B_2$ have the FM state as one of the two lowest-energy states, and the energy difference between these two lowest-energy states is much less than 5 meV/f.u., i.e., 1 meV/atom. It suggests a competition between FM and non-FM ground states in these 4 compounds, which may be easily perturbed by external parameters[44,45].

The AFA ordering (blue plus symbols in Figure 3) has parallel spins within each $T$ layer in the $ab$-plane and alternating antiparallel spins along the $c$-axis. The AAF ordering (blue crosses in Figure 3) has alternating antiparallel spins along the $c$-axis and antiparallel spins within each $T$ layer. These two orderings have opposite-spin sublattices that are connected by a lattice translation and, therefore, are categorized as conventional antiferromagnets[29,30].

The FAA ordering (red diamonds in Figure 3) has ferromagnetically coupled $T$ chains along the $c$-axis and AFM interactions between the nearest chains. $FeNb_2B_2$, $FeMo_2B_2$, and $FeTa_2B_2$ have FAA ground states, which are in good agreement with previous theoretical studies[34,35,42,46]. In addition, $MnMo_2B_2$, $MnRu_2B_2$, $FeZr_2B_2$, $FeRu_2B_2$, $MnW_2B_2$, $MnRe_2B_2$, $FeHf_2B_2$, and $FeW_2B_2$ also have FAA ground states. For this ordering, opposite-spin sublattices are connected only by rotational or mirror symmetries. The Shubnikov point group[47,48] (also known as the spin point group[49]) of the FAA-ordered P4/$mbm$ compound is $^1 4/^1 m^2 m^2 m$, which, according to the classification of Ref.[29], makes it a planar $g$-wave altermagnet.

We obtained a total of 11 altermagnets in this family of compounds, which are also highlighted in bold in Table 1. Seven of them have formation energies within 0.05 eV/atom above the convex hull and are most likely to be synthesizable (purple compositional x-labels in Figure 3a). Except for $FeHf_2B_2$, which has competing altermagnetic and FM ground states, the remaining 10 phases have well-defined altermagnetic ground states. Except for $FeZr_2B_2$ and $FeHf_2B_2$, which have weak magnetic moments, the remaining 9 phases have large magnetic moments of more than 1.3 $\mu_B$ on the magnetic atoms. Overall, there are six thermodynamically and magnetically most promising altermagnetic phases: $FeNb_2B_2$, $FeTa_2B_2$, $FeMo_2B_2$, $FeW_2B_2$, $MnMo_2B_2$, and $MnW_2B_2$.

**Table 1.** Predicted phases within 0.2 eV/atom above the convex hull. Phases are classified according to their structure (tetragonal or orthorhombic), hull distance (within 0.05 or 0.2 eV/atom above the convex hull), and magnetism (nonmagnetic or magnetic). Convex hull phases ($E_d = 0$) are further highlighted by underlining. Altermagnetic phases are further highlighted in bold.

| | | Nonmagnetic | Magnetic |
|---|---|---|---|
| Tetragonal $P4/mbm$ | $E_d \leq 0.05$ (eV/atom) | <u>VNb$_2$B$_2$</u>, <u>VMo$_2$B$_2$</u>, <u>CrNb$_2$B$_2$</u>, <u>CrMo$_2$B$_2$</u>, FeTc$_2$B$_2$, CoTc$_2$B$_2$, <u>NiTc$_2$B$_2$</u>, NiRu$_2$B$_2$, <u>VTa$_2$B$_2$</u>, <u>VW$_2$B$_2$</u>, <u>CrTa$_2$B$_2$</u>, <u>CrW$_2$B$_2$</u>, MnHf$_2$B$_2$ | <u>MnNb$_2$B$_2$</u>, **MnMo$_2$B$_2$**, FeNb$_2$B$_2$, FeMo$_2$B$_2$, CoNb$_2$B$_2$, <u>MnTa$_2$B$_2$</u>, **MnW$_2$B$_2$**, **FeHf$_2$B$_2$**, **FeTa$_2$B$_2$**, **FeW$_2$B$_2$** |
| | $0.05 < E_d \leq 0.2$ (eV/atom) | VTc$_2$B$_2$, VRu$_2$B$_2$, MnZr$_2$B$_2$, CoZr$_2$B$_2$, CoMo$_2$B$_2$, CoRu$_2$B$_2$, NiZr$_2$B$_2$, NiNb$_2$B$_2$, NiMo$_2$B$_2$, NiRh$_2$B$_2$, VRe$_2$B$_2$, FeRe$_2$B$_2$, FeOs$_2$B$_2$, CoHf$_2$B$_2$, CoW$_2$B$_2$, CoRe$_2$B$_2$, CoOs$_2$B$_2$, NiHf$_2$B$_2$, NiTa$_2$B$_2$, NiW$_2$B$_2$, NiRe$_2$B$_2$, NiOs$_2$B$_2$, NiIr$_2$B$_2$ | VZr$_2$B$_2$, CrZr$_2$B$_2$, CrTc$_2$B$_2$, MnTc$_2$B$_2$, **MnRu$_2$B$_2$**, **FeZr$_2$B$_2$**, **FeRu$_2$B$_2$**, FeRh$_2$B$_2$, CoRh$_2$B$_2$, VHf$_2$B$_2$, CrHf$_2$B$_2$, CrRe$_2$B$_2$, **MnRe$_2$B$_2$**, CoTa$_2$B$_2$ |
| Orthorhombic $Immm$ | $E_d \leq 0.05$ (eV/atom) | CrMo$_2$B$_2$, <u>FeMo$_2$B$_2$</u>, FeTc$_2$B$_2$, <u>CoNb$_2$B$_2$</u>, <u>CoMo$_2$B$_2$</u>, <u>CoTc$_2$B$_2$</u>, NiNb$_2$B$_2$, <u>NiMo$_2$B$_2$</u>, NiTc$_2$B$_2$, CrW$_2$B$_2$, FeTa$_2$B$_2$, <u>FeW$_2$B$_2$</u>, FeRe$_2$B$_2$, CoTa$_2$B$_2$, <u>CoW$_2$B$_2$</u>, CoRe$_2$B$_2$, NiTa$_2$B$_2$, <u>NiW$_2$B$_2$</u> | MnNb$_2$B$_2$, MnMo$_2$B$_2$, MnTc$_2$B$_2$, FeNb$_2$B$_2$, MnTa$_2$B$_2$, <u>MnW$_2$B$_2$</u> |
| | $0.05 < E_d \leq 0.2$ (eV/atom) | VZr$_2$B$_2$, VNb$_2$B$_2$, VMo$_2$B$_2$, VTc$_2$B$_2$, CrZr$_2$B$_2$, CrNb$_2$B$_2$, CrTc$_2$B$_2$, MnRu$_2$B$_2$, FeRu$_2$B$_2$, CoZr$_2$B$_2$, CoRu$_2$B$_2$, NiRu$_2$B$_2$, NiRh$_2$B$_2$, NiPd$_2$B$_2$, VHf$_2$B$_2$, VTa$_2$B$_2$, VW$_2$B$_2$, CrHf$_2$B$_2$, CrTa$_2$B$_2$, CrRe$_2$B$_2$, FeLa$_2$B$_2$, FeHf$_2$B$_2$, CoHf$_2$B$_2$, NiRe$_2$B$_2$, NiIr$_2$B$_2$, NiPt$_2$B$_2$ | CrRu$_2$B$_2$, MnZr$_2$B$_2$, MnRh$_2$B$_2$, FeZr$_2$B$_2$, FeRh$_2$B$_2$, CoRh$_2$B$_2$, MnHf$_2$B$_2$, MnRe$_2$B$_2$, FeIr$_2$B$_2$, CoIr$_2$B$_2$ |

To confirm the predicted altermagnetism, we plotted the nonrelativistic spin-polarized band structures for these selected six phases in Figure 4a. Along the chosen lines in reciprocal space, the band structure exhibits momentum-dependent spin splitting. Altermagnetic band splitting in FeNb$_2$B$_2$ and FeTa$_2$B$_2$ is also mentioned in Ref.[50]. The magnitude of spin splitting can be as large as ~0.2–0.3 eV near the Fermi level. This magnitude is relatively small compared to some known altermagnets, such as CrSb and MnTe (splitting ~1.1–1.2 eV)[30]. However, it is comparable to the intermediate-spin-splitting altermagnets, such as KRu$_4$O$_8$ and Mn$_5$Si$_3$ (splitting ~0.15–0.3 eV)[30]. It is significantly larger compared to oxide altermagnets, such as La$_2$CuO$_4$ and LaMnO$_3$, where the small spin splitting of order 0.01 eV[30] appears only due to small TO$_6$ octahedra rotations.

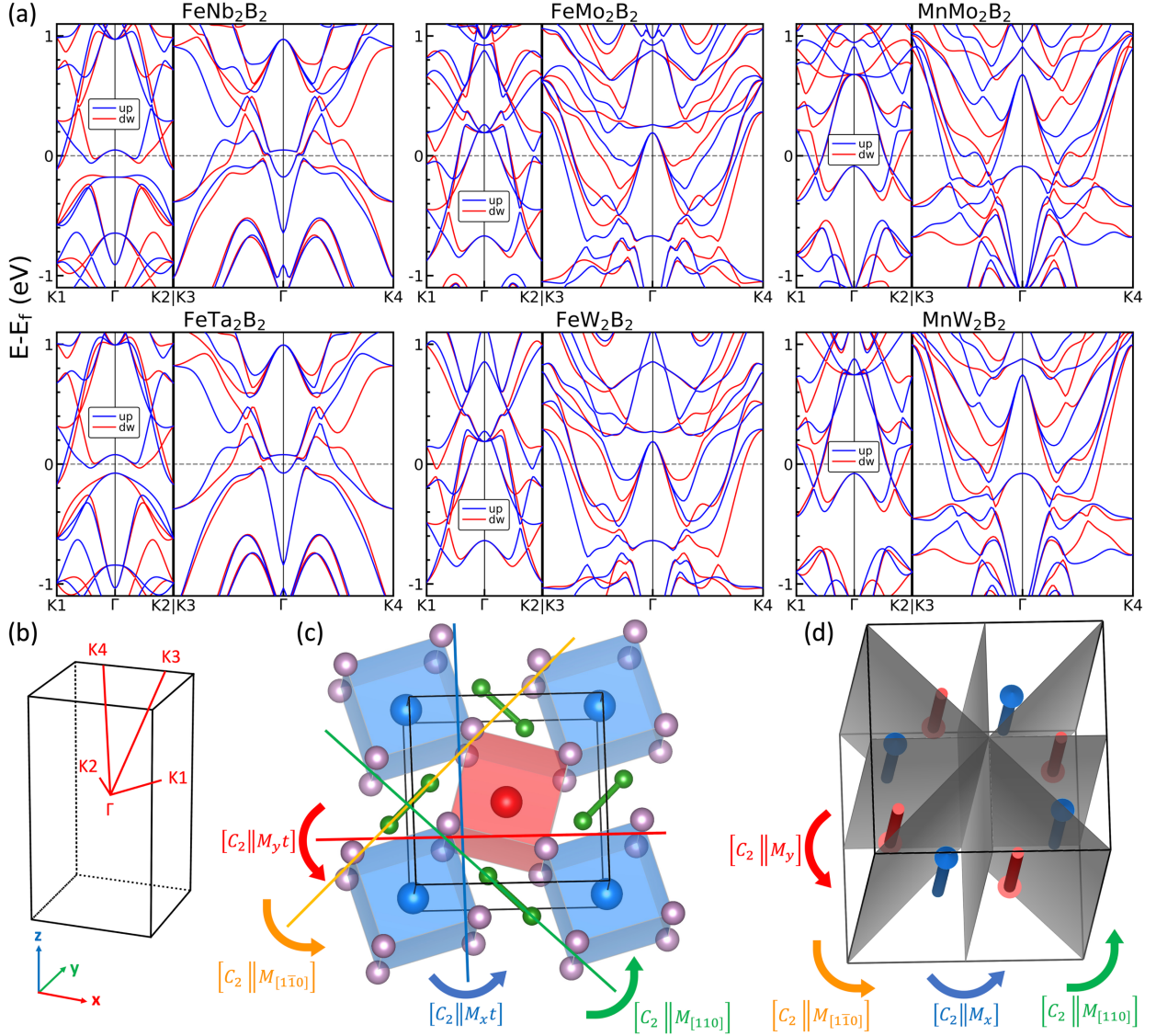

**Figure 4.** (a) Spin-polarized electronic band structure of the magnetic ground state of the tetragonal FeNb$_2$B$_2$, FeTa$_2$B$_2$, FeMo$_2$B$_2$, FeW$_2$B$_2$, MnMo$_2$B$_2$, and MnW$_2$B$_2$. Bands for spin-up and spin-down are shown in blue and red curves, respectively. (b) The k-paths in the Brillouin zone (BZ) of the primitive cell. Apart from Γ(0, 0, 0), non-high-symmetry points are denoted as follows: K1(1/4, 1/2, 0), K2(-1/4, 1/2, 0), K3(1/4, 1/2, 1/2), K4(-1/4, 1/2, 1/2). (c) Crystal structure with two opposite-spin sublattices (blue and red) and possible sublattice-transposing transformations containing four real-space mirror planes. Each mirror plane is represented by a line indicating its projected position in the *xy*-plane. (d) BZ and spin-degenerate nodal planes protected by four momentum-space mirror planes.

To further illustrate how opposite spins compensate each other in both real and momentum spaces, the magnetic structure and the Brillouin zone (BZ) with nodal planes are displayed in Figures 4c and 4d, respectively. When describing the symmetry transformations in altermagnets, a notation within nonrelativistic spin groups is used, $[C_2||A]$. The $C_2$ rotation on the left of the double vertical bar transforms up and down spins into each other in spin space. The $A$ operation on the right of the double vertical bar is a proper rotation ($R$) or mirror plane ($M$) followed by an optional translation ($t$) in real space, which acts simultaneously with the spin-space $C_2$ transformation[29,30,32]. In Figure 4c, the opposite-spin sublattices are highlighted in the same color as those for the opposite spins on the $T$ atoms, respectively. Each $T$ atom is located at the center of a tetragonal prism made of $M$ atoms. For each spin orientation, the magnetic $T$ atom, the coordinated nonmagnetic $M$ atoms, and the crystal field caused by the $M$ atoms together make up the sublattice of that spin. It is the nonmagnetic $M$ atoms that break the space translation and inversion symmetries connecting the opposite-spin sublattices.

On the other hand, some rotational or mirror symmetries connecting the opposite-spin sublattices are present in altermagnets. Four such transformations are indicated in Figure 4c using the notation introduced above. (Due to the existence of the $M_z$ mirror symmetry in this crystal, the indicated four mirror planes can also be replaced by twofold rotation axes at the same projected position, i.e., $[C_2||M_y t] \to [C_2||C_{2x} t]$, $[C_2||M_x t] \to [C_2||C_{2y} t]$, $[C_2||M_{[1\bar{1}0]}] \to [C_2||C_{2[110]}]$, and $[C_2||M_{[110]}] \to [C_2||C_{2[1\bar{1}0]}]$.) Corresponding to the compensation of opposite spins in real space, opposite-spin electronic states in momentum space also compensate each other by the same set of rotational or mirror symmetries (Figures 4a and 4d). In Figure 4d, BZ regions with compensating opposite-spin electronic states are separated by nodal planes. At the nodal planes, spin-up and spin-down electronic states are degenerate, which are protected by the same set of four transformations as those in real space. Away from the nodal planes, opposite-spin electronic bands are connected to each other by these transformations, which are shown in Figure 4a. Hence, despite the electronic band spin splitting, the opposite-spin band compensation in the BZ leads to zero net magnetization, which is characteristic of altermagnetism. Note that in conventional antiferromagnetism, spin-up and spin-down electronic bands are degenerate throughout the BZ.

The spin-polarized DOS for the altermagnetic ground states of these compounds are shown in Figure S3. The Fermi level is located away from any major sharp peaks, suggesting the electronic stability of the magnetic ground states in contrast to the instability of the nonmagnetic states. The dispersive bands seen in Figure 4a suggest that these materials should have metallic properties with reasonably large conductivity.

Altermagnets can have useful properties for applications in spintronic and magnonic devices[51], including the spin-splitter effect[52–55], anomalous transport properties enabled by the broken time-reversal symmetry[56], and chiral magnon splitting[57–60]. In the following sections, we explore these effects for the six boride compounds mentioned in Figure 4.

**Spin-Splitter Effect.** In some altermagnets, due to their spin-split band structure, the bulk charge current is accompanied by a spin current. In the nonrelativistic limit, this spin current is polarized along the altermagnetic order parameter and is described by the symmetric spin-conductivity tensor $\sigma_{ij}^s$. The flow direction of the spin current depends on the crystallographic direction of the charge current. The generation of a transverse spin current, called the spin-splitter (or spin-splitting) effect[52–55], is a promising method to induce magnetization torque in a thin bilayer film consisting of an altermagnet and a ferromagnet. The charge-to-spin conversion efficiency in such bilayers could exceed that of spin-orbit torque bilayers[61], where the transverse spin current is generated thanks to the spin Hall effect. Moreover, in contrast to conventional spin-orbit-torque bilayers, where symmetry forces the polarization of the spin current to lie in the in-plane $\hat{\mathbf{z}} \times \mathbf{E}$ direction ($\hat{\mathbf{z}}$ is normal to the film plane and $\mathbf{E}$ is the electric field), the polarization of the spin current in altermagnets is determined by the orientation of the order parameter. If the latter has an out-of-plane component, the spin-splitter effect could be used to achieve field-free switching of perpendicular magnetization, which is a key requirement for downscalable magnetic memory applications[62].

The spin conductivity tensor $\sigma_{ij}^s$ vanishes in altermagnets belonging to spin Laue groups $^14/^1m^2m^2m$, $^16/^1m^2m^2m$, $^1\bar{3}^2m$, $^26/^2m$, $^26/^2m^2m^1m$, and $^1m^1\bar{3}^2m$[63]. These include the known hexagonal altermagnets MnTe and CrSb (spin Laue group $^26/^2m^2m^1m$) and the boride *P4/mbm* altermagnets studied here (spin Laue group $^14/^1m^2m^2m$). By reducing the symmetry, strain can induce finite spin conductivity (and hence the spin-splitter effect) in such altermagnets[63]. This strain-induced spin conductivity is described by the piezo-spin-galvanic tensor $P_{ijkl}$ defined as follows: $\sigma_{ij}^s = P_{ijkl}\varepsilon_{kl}$. In $^14/^1m^2m^2m$ altermagnets, the non-zero elements of this tensor are $P_{xyxx} = -P_{xyyy}$, $P_{xxxy} = -P_{yyxy}$, and $P_{xzyz} = -P_{yzxz}$[63].

Consider a (100) or (110)-oriented thin film of an altermagnetic *P4/mbm* boride grown epitaxially on a suitable substrate. The relevant shear strain components for such films are $\varepsilon_{xx} - \varepsilon_{yy}$ or $\varepsilon_{xy}$, respectively, with respect to the standard crystallographic reference frame. In this geometry, a charge current flowing perpendicular to the tetragonal axis in the film plane induces a spin current flowing across the thickness of the film, which can be exploited, for example, for spin torque generation in a proximate magnetic film. The charge-to-spin conversion ratio in this geometry is called the spin-splitter angle $\theta_{SS} = \sigma_\perp^s/\sigma_\parallel$, where $\sigma_\parallel = \sigma_{xx}$ is the conductivity measured in the *ab* plane (orthogonal to the tetragonal axis). We define the spin-splitter gauge factor $\xi_{ijkl} = P_{ijkl}/\sigma_\parallel = \theta_{SS}/\varepsilon$, where $\varepsilon$ is the shear strain magnitude. The gauge factors corresponding to (100) and (110)-oriented films are $\xi_{xyxx}$ and $\xi_{xxxy}$, respectively.

We calculate the gauge factors $\xi_{xyxx}$ and $\xi_{xxxy}$ using the first-principles band structures of six altermagnetic *P4/mbm* borides with a 1% applied shear strain of the corresponding geometry. The charge and spin conductivities are calculated using the Boltzmann approximation[64] assuming constant transport relaxation time $\tau$. For the given spin channel (↑ or ↓), the conductivity tensor is

$$\sigma_{ij}^{\uparrow/\downarrow} = \tau \sum_n \int v_{ni}^{\uparrow/\downarrow} v_{nj}^{\uparrow/\downarrow} \frac{\partial f(E_n^{\uparrow/\downarrow})}{\partial \mu} \frac{d^3k}{(2\pi)^3}, \qquad (1)$$

where $E_n^{\uparrow/\downarrow}$ and $\boldsymbol{v}_n^{\uparrow/\downarrow}$ are the energy and group velocity of the *n*-th band for spin ↑ or ↓. The charge and spin conductivities are $\sigma_{ij} = \sigma_{ij}^\uparrow + \sigma_{ij}^\downarrow$ and $\sigma_{ij}^s = \sigma_{ij}^\uparrow - \sigma_{ij}^\downarrow$. Note that $\sigma_{ij}^s$ is odd under time reversal, and its observation requires an unequal population of altermagnetic domains.

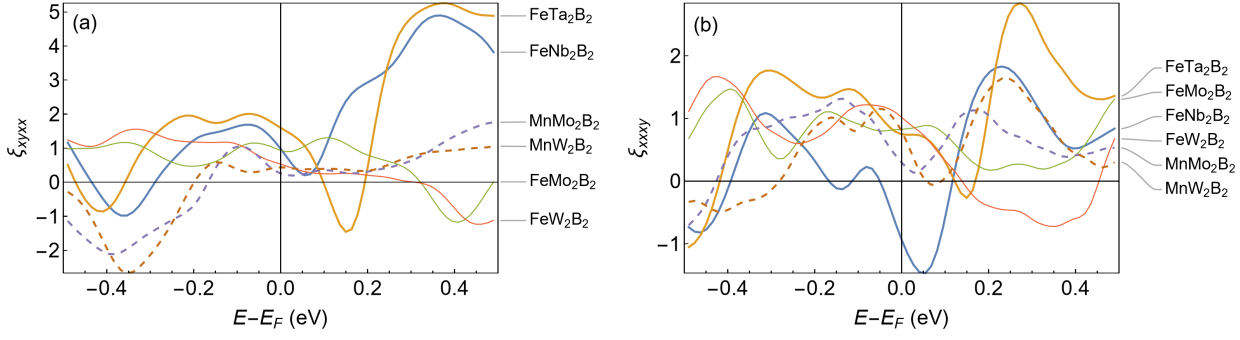

**Figure 5.** Spin-splitter gauge factors (a) $\xi_{xyxx}$ and (b) $\xi_{xxxy}$ as a function of the Fermi energy in six *P4/mbm* altermagnetic borides (see labels). The gauge factors are defined as $\xi_{ijkl} = P_{ijkl}/\sigma_{xx}$ and correspond to the ratio of the spin-splitting angle and the magnitude of the applied strain.

For both strain geometries, the two spins contribute equally to the charge conductivity (which is purely longitudinal) and the spin conductivity (which is purely transverse). Because $\tau$ cancels in the definition of the spin-splitter angle, it is sufficient to calculate the relevant components of the tensor $\sigma_{ij}^{\uparrow}/\tau$, which depend only on the band structure. For this purpose, we use the smooth Fourier interpolation technique[65] implemented in the BOLTZTRAP2 package[66]. The Fermi temperature was set at 300 K to improve the convergence of the BZ integration. The resulting gauge factors $\xi_{xyxx}$ and $\xi_{xxxy}$ are shown in Figure 5 as a function of the Fermi energy.

Figure 5 shows that both spin-splitter gauge factors are of order unity in all six compounds, which is the natural situation for good metals. This means that a typical epitaxial strain of order 1% can induce a spin-splitter angle of order 1% if the film is initialized in a single altermagnetic domain state. While such spin-splitter effect can be observed experimentally, it is small compared to typical spin-Hall sources like Pt[61]. A much larger spin-splitter gauge factor exceeding 30 was predicted for the *p*-doped MnTe altermagnetic semiconductor, where its anomalously large magnitude is due to the degeneracy of the valence band maximum[63].

**Anomalous Hall Effect.** The anomalous Hall effect (AHE) refers to the generation of a transverse voltage in a material carrying an electric current, occurring without the need for an external magnetic field[67]. The AHE cannot exist in nonmagnetic materials and is usually found in ferromagnets. Conventional antiferromagnets cannot sustain the AHE due to the lack of required time-reversal-symmetry breaking. Altermagnets provide a new opportunity for realizing large AHE without a large net magnetization[56]. In spintronic applications, AHE can be used to read out the altermagnetic domain state.

To examine the AHE for metallic altermagnets in this study, we computed the anomalous Hall conductivity (AHC) using the Kubo formula. We employed maximally localized Wannier functions (MLWFs) to construct an effective low-energy model. The Wannierization procedure was performed using the Wannier90 code[68,69] based on the Bloch wavefunctions obtained from VASP with spin-orbit coupling included. Atomic orbital-like MLWFs corresponding to Fe/Mn-*d*, Nb/Ta/Mo/W-*d*, and B-*p* states were used as initial projections. The resulting tight-binding

Hamiltonian accurately reproduced the DFT band structure within an energy window of approximately ±1 eV around the Fermi level.

The Berry curvature $\hat{\Omega}_n(k)$ was calculated by the Kubo formula[70],

$$\Omega_n^{ij}(k) = -\hbar^2 \sum_{m \neq n} \frac{2 \operatorname{Im}[\langle u_{nk}|\hat{v}_i|u_{mk}\rangle\langle u_{mk}|\hat{v}_j|u_{nk}\rangle]}{(\epsilon_{nk}-\epsilon_{mk})^2}, \qquad (2)$$

where $u_{n\mathbf{k}}$ and $\epsilon_{n\mathbf{k}}$ are the eigenstates and eigenvalues of the Hamiltonian $\hat{H}(\mathbf{k})$, and $\hat{v}_i = \frac{1}{\hbar}\frac{\partial \hat{H}(k)}{\partial k_i}$ is the velocity operator. The AHC $\sigma_A^{ij}$ was computed by integrating the Berry curvature over the occupied states in the BZ,

$$\sigma_A^{ij} = -\hbar e^2 \sum_n \int_{BZ} \frac{d^3k}{(2\pi)^3} f_n(k)\Omega_n^{ij}(k), \qquad (3)$$

where $f_n(\mathbf{k})$ is the Fermi-Dirac distribution function. This calculation was performed using the WannierTools package[71] utilizing a dense $k$-mesh of $200 \times 200 \times 200$, which was found to yield well-converged results. A constant broadening parameter of 10 meV was used to stabilize the Berry curvature integration over the BZ.

Our magnetic anisotropy calculations (see Text S1) reveal that FeNb$_2$B$_2$ and FeTa$_2$B$_2$ possess an intrinsic easy axis lying within the xy-plane, consistent with previous reports[46]. We also reveal that MnMo$_2$B$_2$ and MnW$_2$B$_2$ have an in-plane spin alignment, while FeMo$_2$B$_2$ and FeW$_2$B$_2$ have an easy magnetization axis along the out-of-plane direction. Thus, among the six borides considered, two (FeMo$_2$B$_2$ and FeW$_2$B$_2$) are easy-axis, and four (FeNb$_2$B$_2$, FeTa$_2$B$_2$, MnMo$_2$B$_2$, and MnW$_2$B$_2$) are easy-plane. The calculated magnetic anisotropy energies (MAE), MAE = $E_{[100]} - E_{[001]}$, are listed in Table 2. The relatively large MAE observed for FeTa$_2$B$_2$ and FeW$_2$B$_2$ can be attributed to the stronger spin-orbit coupling associated with the heavy 5$d$ elements Ta and W. Surprisingly, MnW$_2$B$_2$ exhibits a much smaller MAE despite containing W. This reduction is likely due to the half-filled 3$d^5$ configuration of Mn, which leads to quenched orbital moments, thereby diminishing the spin-orbit-driven anisotropy compared to Fe-based borides.

Symmetry analysis shows that the AHC tensor vanishes if the Néel vector is oriented along the [001] direction. In contrast, a finite component $\sigma_A^{zx}$ is allowed for the [100] orientation. We calculate this component for the [100] orientation in all the above six compounds, even though it is not always the ground state. Easy-axis compounds (FeMo$_2$B$_2$ and FeW$_2$B$_2$) could be switched to the in-plane configuration by applying a magnetic field along the easy-axis [001] direction, inducing the spin-flop transition. Figure S4 presents the calculated AHC as a function of the Fermi energy for the [100] orientation, and Table 2 lists the AHC values at the Fermi level. For all compounds, the AHC exhibits large variations as a function of the Fermi energy within the shown ±0.5 eV window around the Fermi level.

Easy-plane FeNb$_2$B$_2$, FeTa$_2$B$_2$, and MnMo$_2$B$_2$ compounds exhibit large AHC of 57, 81, and -34 S/cm, respectively, in their ground state. An earlier calculation[46] obtained $|\sigma_A^{zx}|$ ~50–100 S/cm for both FeNb$_2$B$_2$ and FeTa$_2$B$_2$ with the same generalized gradient approximation (GGA) functional, which is in reasonable agreement with our results. FeMo$_2$B$_2$ and FeW$_2$B$_2$ also exhibit large AHC in the in-plane configuration, -35 and 300 S/cm, respectively, which could be obtained in the spin-flop state. The large AHC in this family of altermagnets make them potentially one of the pioneering classes of compensated collinear magnets exhibiting intrinsic AHE.

**Table 2.** Magnetic anisotropy energy (meV per magnetic atom) and anomalous Hall conductivity $\sigma_A^{zx}$ (S/cm) at the Fermi level for **L** ∥ [100]. The values of AHC for easy-plane compounds are highlighted in bold.

| Compound | MAE | AHC |
|---|---|---|
| FeNb$_2$B$_2$ | -0.32 | **57** |
| FeTa$_2$B$_2$ | -2.07 | **81** |
| FeMo$_2$B$_2$ | 0.34 | -35 |
| FeW$_2$B$_2$ | 1.67 | 300 |
| MnMo$_2$B$_2$ | -0.27 | **-34** |
| MnW$_2$B$_2$ | -0.15 | **0** |

**Chiral Magnons.** Next, we focus on the magnonic properties. Magnons, also known as spin waves, are collective spin excitations in magnetically ordered crystals, which are capable of transporting angular momentum over long distances. In general, in metallic systems, the lifetimes of high-frequency magnons are significantly reduced due to their strong coupling to the Stoner continuum. Nevertheless, long-lived magnons have been found in some metallic antiferromagnets[72]. Magnons in altermagnets exhibit chiral splitting throughout the Brillouin zone, except for the nodal planes and lines, similarly to the spin splitting of the electronic bands[57–60].

To calculate the exchange coupling interactions and spin wave spectrum, we use the Heisenberg model approach with the spin-polarized version of RKKY exchange interaction parameters[73–76]. We developed the corresponding computational program (see Text S1) in DFT using a full-potential linearized augmented plane wave method with the FlapwMBPT[77,78] code. In general, the spin-polarized RKKY (or long wavelength[79]) approximation is better suited for relatively localized moment systems and magnetic insulators, while for more itinerant metallic systems the error can be above 20% for the nearest neighbors exchange coupling parameter (see discussion in Ref.[79]).

Our obtained real-space exchange coupling parameters, $J_{ij}$, for the first six nearest *T-T* neighbors in FeNb$_2$B$_2$, FeTa$_2$B$_2$, and FeMo$_2$B$_2$ are provided in Table 3 and indicated alongside the magnetic structure in Figure 6d. $J_{ij}$ are defined by the effective Heisenberg model, $E = -\sum_{ij} J_{ij} \hat{m}_i \hat{m}_j$ (see discussion near Eq. (27) in Ref.[74]), where $\hat{m}_i$ are unit vectors, and each atomic pair is counted twice. The FM intrachain coupling ($J_l$) clearly dominates. The results obtained using the local density approximation (LDA) and the GGA are very similar. The total exchange coupling between a given site and the rest of the crystal is given by $J_0 = \sum_j J_{ij} \hat{m}_i \hat{m}_j$. The Néel temperature ($T_N$) is estimated by $T_N = (2/3) J_0 / k_B$ within the mean-field approximation (MFA).

**Table 3.** Exchange parameters $J_{ij}$ (meV) in FeNb$_2$B$_2$, FeTa$_2$B$_2$, and FeMo$_2$B$_2$ by GGA and LDA. Both results were obtained in GGA-optimized geometry. $N_j$: number of neighbors of a given type. $R_{ij}$ (Å): distance to the given neighbor. Antiparallel pairs are highlighted in bold. $T_N$ (K): Néel temperature.

|  |  | FeNb$_2$B$_2$ | | | FeTa$_2$B$_2$ | | | FeMo$_2$B$_2$ | | |
|---|---|---|---|---|---|---|---|---|---|---|
| Index of pair | $N_j$ | $R_{ij}$ | $J_{ij}$ (GGA) | $J_{ij}$ (LDA) | $R_{ij}$ | $J_{ij}$ (GGA) | $J_{ij}$ (LDA) | $R_{ij}$ | $J_{ij}$ (GGA) | $J_{ij}$ (LDA) |
| 1 | 2 | 3.28 | 11.7 | 11.39 | 3.29 | 11.32 | 11.78 | 3.15 | 5.32 | 5.51 |
| 2 | 4 | 4.17 | **0.06** | **0.40** | 4.15 | **1.22** | **1.21** | 4.08 | **0.70** | **0.91** |
| 3 | 8 | 5.30 | **-1.93** | **-1.72** | 5.30 | **-1.83** | **-1.82** | 5.16 | **-1.66** | **-1.05** |
| 4 | 4 | 5.89 | 1.95 | 1.97 | 5.87 | 2.60 | 2.30 | 5.77 | 0.62 | 0.55 |
| 5 | 2 | 6.57 | -2.09 | -2.60 | 6.57 | -2.70 | -2.90 | 6.31 | -0.28 | -0.52 |
| 6 | 8 | 6.74 | -0.87 | -0.82 | 6.73 | -0.67 | -0.79 | 6.58 | 0.36 | 0.32 |
| $J_0$ |  |  | 35.17 | 31.07 |  | 32.05 | 30.39 |  | 25.92 | 19.53 |
| $T_N$ |  |  | 272 | 240 |  | 248 | 235 |  | 200 | 151 |

Ref.[46] reported $J_1$ to $J_3$ for FeNb$_2$B$_2$ and FeTa$_2$B$_2$ based on the total energy analysis of the four magnetic configurations, which are larger than ours in magnitude. Using these parameters, the $T_N$ reported by Ref.[46] by classical Monte Carlo simulations are higher than room temperature, while our $T_N$ are below room temperature. If estimated by the MFA using their parameters, their $T_N$ will be approximately twice as large as ours. The larger $J_{ij}$ and $T_N$ reported by Ref.[46] compared to this study might be attributed to the following factors: (i) Local magnetic moments obtained by GGA+U calculations in Ref.[46] are ~3 $\mu_B$, while our magnetic moments obtained by GGA are ~1.8–1.9 $\mu_B$ (see Figure 3a). (ii) The mapping of all magnetic interactions to the first three nearest neighbors in Ref.[46] can be inappropriate, e.g., $J_4$ and $J_5$ are nonnegligible (see Table 3).

The magnonic bands in altermagnets, similar to the electronic bands, exhibit lifted degeneracy for bands of opposite magnon chiralities. Such chiral magnon splitting has been reported for MnTe[58] and CrSb[80]. Spin wave dispersion for FeNb$_2$B$_2$ is shown in Figure 6a. The chiral magnon splitting in FeNb$_2$B$_2$ can reach as large as ~10 meV. This splitting is a few times larger compared to MnTe[58] and a few times smaller compared to the predictions for CrSb[59]. The splitting in FeMo$_2$B$_2$ and FeTa$_2$B$_2$ is relatively smaller, reaching as large as ~5 meV. Opposite-chirality magnons swap under the same symmetry operations as the bands for opposite-spin electrons (see Figure 4d). The dispersion near the Γ point is linear, as in all collinear antiferromagnets.

To track the origin of the magnonic band splitting in the magnetic exchange coupling in the Heisenberg model, we recalculated spin wave dispersions by truncating the exchange interactions in FeNb$_2$B$_2$ at $J_{23}$ or $J_{25}$, as displayed in Figures 6b and 6c, respectively. The degeneracy is not lifted by interactions up to $J_{23}$, which corresponds to the distance $R \sim 13$ Å. Including two more coordination spheres, $J_{24}$ and $J_{25}$, leads to the chiral splitting. Figure 6c approximates Figure 6a well, and including further interactions beyond $J_{25}$ has a minor effect on the dispersion. This suggests that the chiral splitting is primarily determined by $J_{24}$ and $J_{25}$.

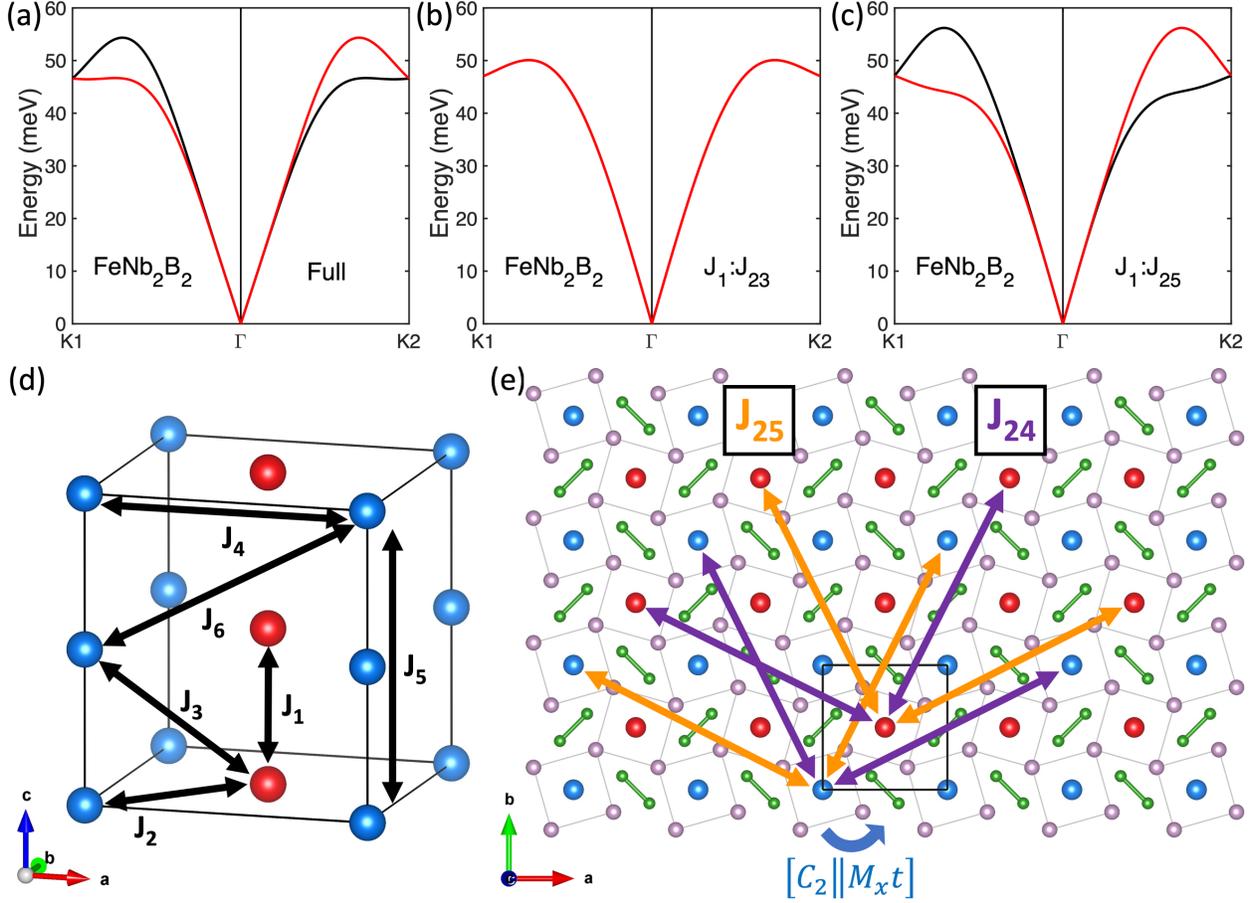

**Figure 6.** (a) Full spin-wave spectrum of the tetragonal FeNb$_2$B$_2$ and spin-wave spectra obtained from real-space exchange coupling parameters (b) $J_1$ to $J_{23}$ and (c) $J_1$ to $J_{25}$ by GGA. Red and black curves indicate opposite magnon chiralities. Non-high-symmetry points are denoted as follows: K1(1/4, 1/2, 0), K2(-1/4, 1/2, 0). (d) 1 × 1 × 2 supercell with only *T* atoms shown. Blue and red spheres indicate *T* atoms with opposite spins. Black double-headed arrows show the exchange couplings between the first six nearest neighbors of *T* atoms. (e) Purple and orange double-headed arrows show $J_{24}$ and $J_{25}$, respectively.

$J_{24}$ = 0.484 meV and $J_{25}$ = -0.467 meV are two in-plane couplings with identical distances, which are shown in Figure 6e on top of the magnetic structure, including the nonmagnetic atoms. The nonmagnetic atoms act as a medium to transfer the magnetic exchange coupling between the *T* atoms. It is seen that $J_{24}$ and $J_{25}$ are distinct due to the asymmetric environments created by the nonmagnetic atoms, which is the origin of the band splitting. It is easy to check that these are the shortest such couplings. In Figure 6e, the magnetic interaction environments of the two opposite spins in the reference unit cell are also portrayed. They can also be transformed into each other by the four symmetry operations indicated in Figure 4c. (Only one of them is shown in Figure 6e for clarity.) This explains why the symmetry of the magnonic bands with opposite chiralities is as shown in Figure 4d. The spin-wave spectrum for FeNb$_2$B$_2$ out of the $k_z$ plane and those for FeMo$_2$B$_2$ and FeTa$_2$B$_2$ are provided in Figure S5.

**Experimental Synthesis:** We aimed to synthesize the low-energy compositions predicted by our theory, especially the altermagnets in the tetragonal structure. Preliminary experimental observations for the tetragonal FeMo$_2$B$_2$[26], FeNb$_2$B$_2$[27], and FeTa$_2$B$_2$[28] were mentioned a long time ago. In these early papers, methods to produce single-phase samples and information on the magnetic impurities were unavailable. To the best of our knowledge, except for FeMo$_2$B$_2$, no recent synthesis is available for any of the tetragonal $TM_2$B$_2$ phases. There are recent reports on the synthesis of FeMo$_2$B$_2$[43,81]. However, information on the magnetic impurities and whether the samples are suitable for magnetic characterization was unknown there.

To synthesize the $TM_2$B$_2$ ternary borides, synthetic methods that utilize high temperatures, high pressures, and prolonged reaction times have been recently reported for the tetragonal FeMo$_2$B$_2$[43,81], and orthorhombic CoMo$_2$B$_2$ and NiMo$_2$B$_2$[81,82]. The reason for high energy input is that elemental boron (melting point: 2075 °C) and binary boride precursors used for syntheses are highly inert refractory materials that require high energy input to overcome kinetic activation barriers. Alternate sources of boron with higher reactivity, such as NaBH$_4$[83,84] and BI$_3$[85], have been used to address this issue. These methods require strict regulation of the preparation conditions, as these precursors are unstable and sensitive to ambient conditions. In this work, we were inspired by using elemental iodine as a transport agent for crystal growth in solid-state synthesis[86–88]. We investigated the applicability of solid iodine as an activation agent. As a result, an iodine-assisted solid-state synthesis scheme that yields crystalline samples has been developed. When carefully optimized for each system, this facile route can produce high-purity samples below 1100 °C within 24 hours.

Previously, Kovnir *et al*. reported the successful use of BI$_3$ for synthesizing binary transition metal borides and their solid solutions[85]. However, BI$_3$ is an air-, moisture-, and light-sensitive precursor that must be handled in regulated conditions, preferably in an argon-filled glovebox. We aimed to develop an alternative method to obtain clean and crystalline products when the reactions were prepared under ambient conditions with relatively stable starting materials. For instance, for FeMo$_2$B$_2$, attempts to react elemental 2Mo + Fe + 2B did not produce successful results due to the high melting points of the starting materials. When the elements were atomically mixed through arc-melting, it produced a mixture of binary borides.

Therefore, we improved the BI$_3$ method by synthesizing BI$_3$ *in situ* using elemental boron and iodine, which can be handled under ambient conditions. We used elemental precursors with iodine crystals in stoichiometric ratios, for instance, FeMo$_2$B$_2$: 1Fe + 2Mo + 2B + 3I$_2$, maintaining the 1:3 boron to iodine ratio to favor BI$_3$ intermediate. This reaction yielded a mixture of binary molybdenum borides, hinting that iodine is consuming elemental iron. Over time, after conducting dozens of reactions on several $TM_2$B$_2$ ($T$ = Mn, Fe, Co, Ni; $M$ = Nb, Mo, Ta, W) systems, it was established that iodine reacts with the 3$d$ transition metal to form the major iodine-containing byproduct $T$I$_2$. For example, the more I$_2$ is present in the reaction medium, the more Fe is consumed to form FeI$_2$, which is washed away by acid treatment, resulting in binary molybdenum borides as exclusive products. Nonetheless, we cannot exclude the formation of 4$d$/5$d$ metals and boron-containing iodides, e.g., MoI$_3$ and BI$_3$, in smaller quantities, which are water soluble. Thus, the optimization of the nominal composition containing an excess of 3$d$ metal and iodine was performed to minimize the formation of binary or ternary impurity phases. Hence, the amount of I$_2$ added to the reaction ampoule was carefully optimized.

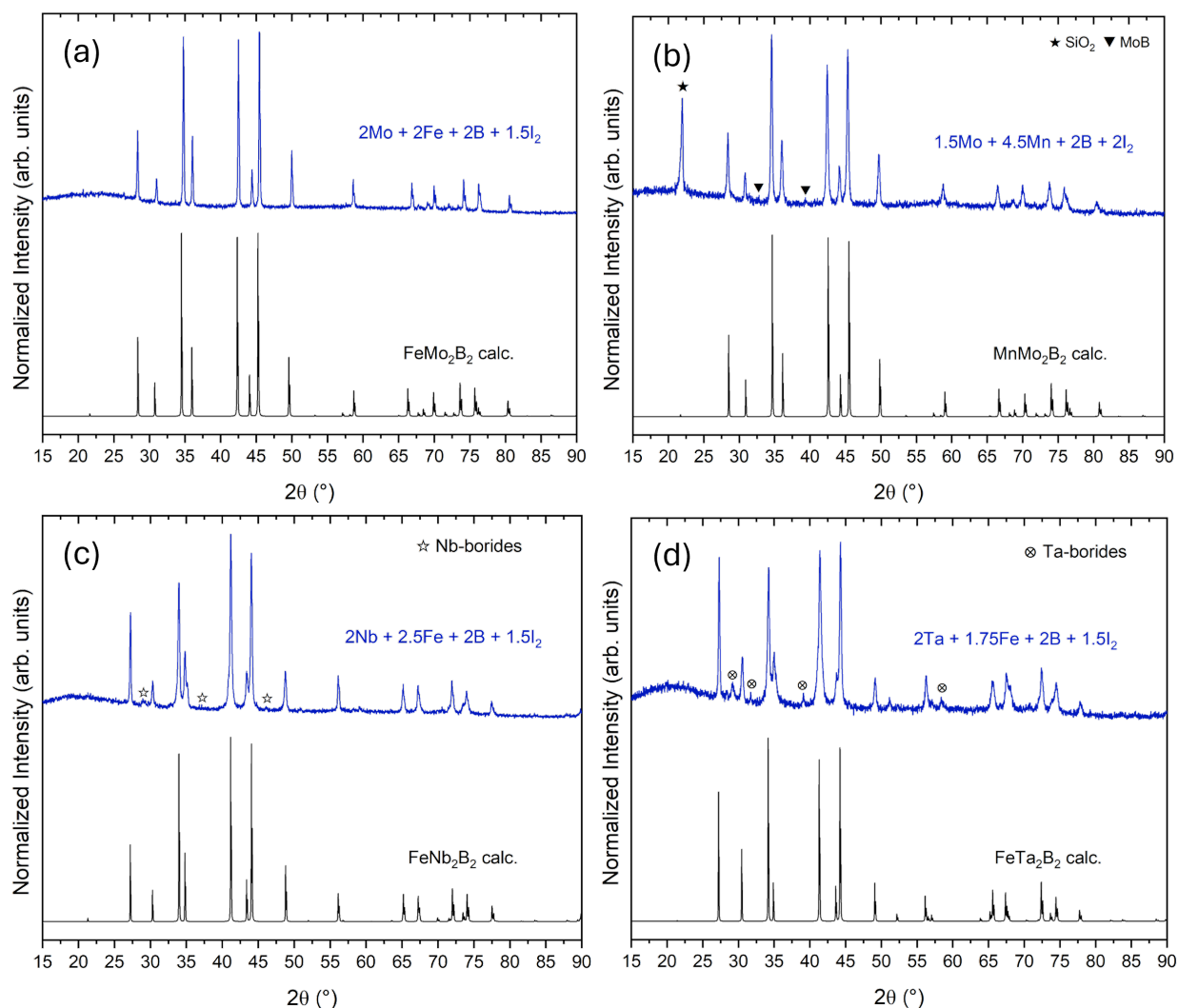

**Figure 7.** Experimental and theoretical PXRD patterns ($\lambda$ = 1.5406 Å) of tetragonal (*P4/mbm*) (a) FeMo$_2$B$_2$, (b) MnMo$_2$B$_2$, (c) FeNb$_2$B$_2$, and (d) FeTa$_2$B$_2$ phases synthesized by the iodine-assisted synthesis method. Optimized loading ratios are shown in blue text.

With this knowledge, a balanced chemical reaction could be written for the FeMo$_2$B$_2$ synthesis reaction as follows: 2Fe + 2Mo + 2B + I$_2$ → FeMo$_2$B$_2$ + FeI$_2$. In general, for *TM*$_2$B$_2$, the required amounts of iodine to minimize admixtures were phase-specific and non-stoichiometric. For 0.1 g scale reactions, the amount of added iodine varied from 40–100 mg. To account for partial iodine sublimation during evacuation and sealing, a 10% excess of iodine was loaded. Cooling the ampoules with liquid $N_2$ or in an ice bath before sealing may help minimize $I_2$ sublimation. The amount of respective $3d$ metal loaded should also be optimized with respect to $I_2$. Loading excessive amounts of $3d$ metal would result in magnetic admixtures, which could interfere with magnetic property measurements.

Using the synthesis method detailed in Text S2, we experimentally synthesized seven ternary borides in $TM_2B_2$ ($T$ = Mn, Fe, Co, Ni; $M$ = Nb, Mo, Ta, W). Four phases, $FeMo_2B_2$, $MnMo_2B_2$, $FeNb_2B_2$, and $FeTa_2B_2$, were found to crystallize in the tetragonal $P4/mbm$ structure. Three phases, $CoMo_2B_2$, $NiMo_2B_2$, and $FeW_2B_2$, were found to crystallize in the orthorhombic *Immm* structure. Six out of seven synthesized phases agree well with the theoretically predicted thermodynamic stability, i.e., the one with a relatively lower formation energy. The only exception is $FeMo_2B_2$, for which the orthorhombic structure was predicted to have a lower formation energy, whereas the tetragonal structure was synthesized. This is not particularly surprising since the tetragonal $FeMo_2B_2$ is merely 0.018 eV/atom higher in the formation energy at zero temperature. The newly developed iodine-assisted solid-state method enables access to four altermagnetic borides.

The experimental and calculated PXRD patterns of the four tetragonal phases and the three orthorhombic phases are provided in Figures 7 and 8, respectively. Detailed optimized reaction conditions and products for all the synthesized phases are provided in Table S1 of the Supporting Information. $FeMo_2B_2$ (Figure 7a), $CoMo_2B_2$ (Figure 8a), and $FeW_2B_2$ (Figure 8b) were synthesized with seemingly no admixtures. For some systems, admixtures that in-lab PXRD did not detect were detected by energy dispersion X-ray spectroscopy (EDX) and high-resolution synchrotron PXRD. For $MnMo_2B_2$ (Figure 7b) synthesis, the reaction of Mn with the silica tube led to cristobalite ($SiO_2$) formation at about 10-30% w/w. The $FeNb_2B_2$ (Figure 7c) sample had traces of niobium borides and occasional elemental Nb. In the case of $FeTa_2B_2$ (Figure 7d), the formation of tantalum oxides and tantalum borides was observed. Carbonization was attempted to prevent the elements from reacting with the silica ampoule. However, syntheses carried out in carbonized tubes resulted in more stable transition metal carbides instead of targeted borides. The $NiMo_2B_2$ (Figure 8c) sample contained trace amounts of nickel borides and, in some cases, elemental Ni. $NiMo_2B_2$, $FeNb_2B_2$, and $FeTa_2B_2$ form at a relatively higher temperature of 1100 °C (Table S1). These three phases are kinetically stable compounds that can be accessed in a short temperature window; hence, their reaction profiles are different from those of others. However, it was difficult to obtain single-phase samples (Table S1). On many occasions, the products were accompanied by minor binary or competing ternary borides. Therefore, fine-tuning the synthesis conditions was necessary.

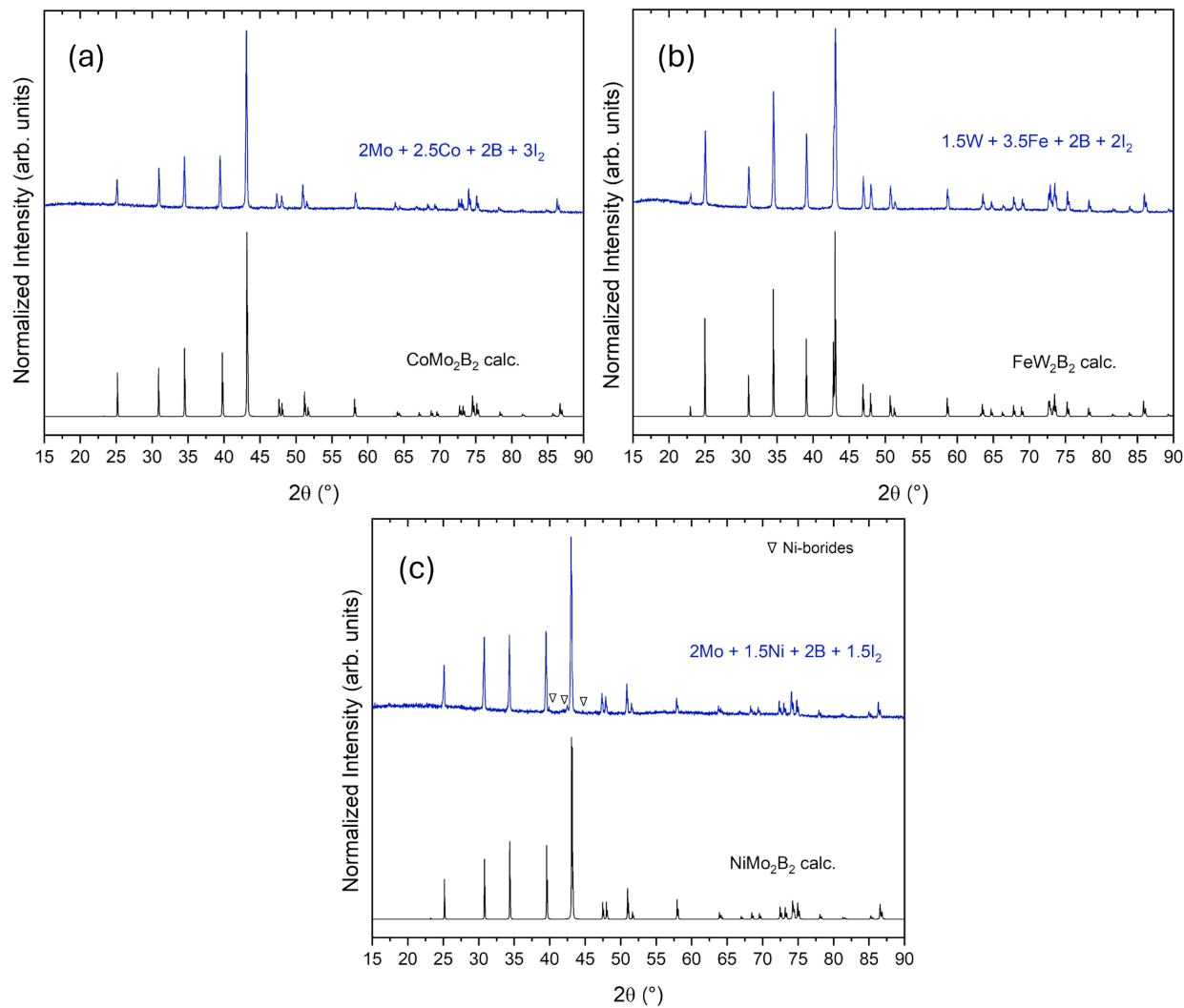

**Figure 8.** Experimental and theoretical PXRD patterns ($\lambda = 1.5406$ Å) of orthorhombic (*Immm*) (a) $CoMo_2B_2$, (b) $FeW_2B_2$, and (c) $NiMo_2B_2$ phases synthesized by the iodine-assisted synthesis method. Optimized loading ratios are shown in blue text.

**Variable-Temperature *In-Situ* Studies:** To shed light on reaction mechanisms, formation temperatures, phase transformations, melting points, etc., of the ternary phases, *in-situ* X-ray diffraction experiments were conducted at the NSLS-II at BNL. For $FeMo_2B_2$, powders of elements were mixed in the $2Mo + 3Fe + 2B$ ratio and were loaded into the capillary sandwiched between $I_2$ crystals (Figure S6). PXRD patterns were recorded every minute as the experiment progressed, and the capillary was subjected to the set heating/cooling profile (Figure 9). At room temperature, peaks corresponding to the starting materials, elemental Mo, Fe, and $I_2$, were visible. Amorphous boron did not show any diffraction peaks. As the temperature gradually increased at a 15 °C per minute rate, iodine quickly sublimed or melted, as evidenced by the complete disappearance of $I_2$ diffraction peaks above 175 °C. Afterward, until 225 °C, only peaks for elemental Mo and Fe were visible. Above 225 °C, $FeI_2$ (melting point 587 °C) formed and remained up to approximately 640 °C. Upon further heating beyond 640 °C, an unidentified phase emerged and coexisted with Mo and Fe until the temperature reached 1000 °C. The presence of liquid $FeI_2$ may serve as a reactive flux, shortening diffusion pathways and promoting component reactivities[89–91].

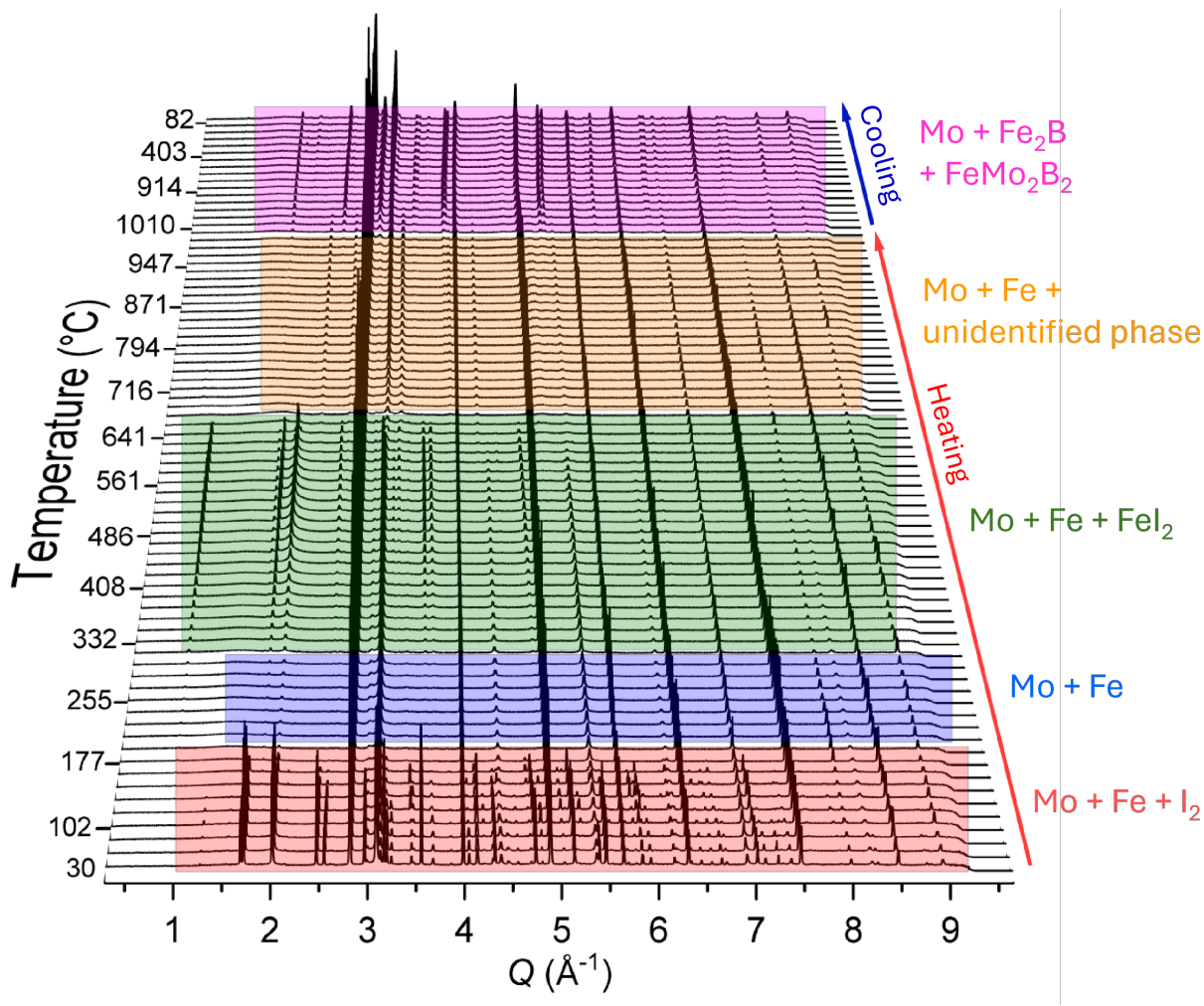

**Figure 9.** 3D waterfall plot of the PXRD patterns collected during *in-situ* experiments conducted at the beamline 28-ID of the NSLS-II at BNL. Reference PXRD patterns are provided in Figure S7.

Meanwhile, peaks corresponding to $Fe_2B$ appeared above 900 °C, followed by minor quantities of the target phase $FeMo_2B_2$. Tetragonal $FeMo_2B_2$ is a thermodynamically stable phase formed only when the temperatures reach ~1000 °C. It remained in the mixture with Mo and $Fe_2B$ after cooling to room temperature. Intense peaks for elemental Mo persisted throughout the experiment during heating and cooling because Mo is a strong X-ray scatterer. The kinetics of the experiment, short reaction time at maximum temperature, concentration differences, and restrictions in the diffusion of precursors may have led to an incomplete reaction that resulted in binary $Fe_2B$ accompanying the $FeMo_2B_2$ phase in the *in-situ* experiments.

**Structure Refinement Using HR-PXRD Data:** Structure refinement was performed using high-resolution synchrotron powder X-ray diffraction (HR-PXRD) data. Rietveld refinement was employed to refine the unit cells and assess mixed occupancy of atomic sites. Note that our allocated beamtime arrived before we fully optimized the synthesis; hence, the samples studied at the synchrotron are not necessarily single-phase and contain considerable amounts of admixtures. Our optimized syntheses, shown in Table S1, produced substantially lower amounts of impurities. Please refer to it for the information on phase purity.

Of the seven ternary boride phases studied, only the four tetragonal phases exhibited mixed occupancy between the $3d/4d$ or $3d/5d$ transition metal sites (Table 4). In the case of $FeMo_2B_2$, the $2a$ Fe site contained 5% Mo (Figure S8a). For $MnMo_2B_2$, there is a higher amount of Mo on the Mn site, 19% (Figure S8b). For $FeNb_2B_2$, no Nb was found on the Fe site; instead, the $4h$ Nb site contained 10% Fe (Figure S8c). Finally, for $FeTa_2B_2$, no Ta/Fe mixed occupancy was detected; instead, 6% Fe site vacancy was observed (Figure S8d). This result was verified by the EDX composition as well (Table 4). Rietveld refinements also indicated a partial site disorder related to the transition metal sites.

$FeMo_2B_2$ was accompanied by ~7% of FeB and MoB binary phases. The $MnMo_2B_2$ sample contained admixture phases that were challenging to identify. Efforts to incorporate known admixtures into the refinement were unsuccessful, likely due to changes in the unit cell dimensions caused by atomic mixing. As a result, only the primary phase was refined. The $FeNb_2B_2$ sample included competing ternary boride phases that have not yet been thoroughly analyzed or identified, and therefore, they were excluded from the refinement. The $FeTa_2B_2$ sample contained significant amounts of TaB.

The orthorhombic phases displayed no signs of mixed occupancies of the transition metal sites (Table 4). The orthorhombic samples still contained binary admixtures. The Rietveld refinement of the $CoMo_2B_2$ phase revealed 8% of CoB (Figure S9a). For $FeW_2B_2$, the composition included 39% of the FeB phase (Figure S9b). For $NiMo_2B_2$, the sample showed >95% phase purity, with a trace amount of nickel borides that could not be identified with certainty (Figure S9c).

The realistic experimental conditions prompted us to improve the theoretical prediction models for these altermagnets to account for mixed occupancy and partial site disorder. Since the first-principles disordered supercell models are computationally very expensive, we only conducted for the representative $FeMo_2B_2$.

$FeMo_2B_2$ was observed to have a fully occupied Mo site and 5% Mo on the Fe site (Table 4). Since no additional magnetic site needs to be added, the four magnetic orderings for the fully ordered phase (Figure 3b) can be easily mapped to a disordered supercell phase with certain magnetic sites occupied by nonmagnetic atoms. We constructed a disordered supercell model using the in-house code as follows. A 160-atom $2 \times 2 \times 4$ supercell of $FeMo_2B_2$ containing 32 Fe atoms was used (Figure S10a). Two of the Fe atoms were substituted by Mo atoms, which led to a 6.25% Mo impurity on the Fe site to approximate the experimental sample. A smaller supercell with one Fe substitution might introduce artificial effects of chemical short-range order and, thus, was not used. All possible combinations for the substitution were generated, and redundant supercell configurations were removed by symmetry only to keep inequivalent configurations. The redundancy was recorded as the weight of each configuration. Then, DFT magnetic calculations were performed for the inequivalent supercell configurations. The final energy for each magnetic ordering was obtained as the weighted average of the configurations' energies (Figure S10b).

**Table 4.** Details of the Rietveld refinements performed on the synchrotron HR-PXRD patterns of $TM_2B_2$ phases obtained at the beamline BL2-1 ($\lambda$ = 0.72964 Å) of SLAC National Accelerator Laboratory, Stanford Synchrotron Radiation Light Source (SSRL). Mixed occupancies of metal atomic sites were only observed in tetragonal phases. Certain admixtures of competing binary or ternary borides with unclear compositions due to atomic mixing were excluded from the refinements. In these cases, the percentage composition of the major phase was not determined. The $T$:$M$ metal ratio determined by EDX is also provided.

| | Boride | % $R_w$ for the model without/with mixed occupancies | Composition derived from the refinement | EDX composition excluding B | Admixtures and phase fractions |
|---|---|---|---|---|---|
| **Tetragonal** $P4/mbm$ | $FeMo_2B_2$ | 11.3/10.3 | $[Fe_{0.95(1)}Mo_{0.05(1)}]Mo_2B_2$ | $Fe_{1.00(3)}Mo_{2.00(4)}$ | FeB 6.5%, MoB 0.5% |
| | $MnMo_2B_2$ | 8.58/8.17 | $[Mn_{0.81(1)}Mo_{0.19(1)}]Mo_2B_2$ | $Mn_{0.99(1)}Mo_{2.01(1)}$ | Binary molybdenum borides |
| | $FeNb_2B_2$ | 13.5/13.2 | $Fe[Nb_{0.90(1)}Fe_{0.10(1)}]_2B_2$ | $Fe_{1.10(7)}Nb_{1.90(7)}$ | Binary niobium borides |
| | $FeTa_2B_2$ | 5.44/5.36 | $Fe_{0.94(1)}Ta_2B_2$ | $Fe_{0.90(2)}Ta_{2.00(2)}$ | TaB 46% |
| **Orthorhombic** $Immm$ | $CoMo_2B_2$ | 6.93 | $CoMo_2B_2$ | $Co_{1.07(3)}Mo_{1.93(3)}$ | CoB 8% |
| | $FeW_2B_2$ | 2.95 | $FeW_2B_2$ | $Fe_{1.03(7)}W_{1.97(7)}$ | FeB 39% |
| | $NiMo_2B_2$ | 6.98 | $NiMo_2B_2$ | $Ni_{1.00(4)}Mo_{2.00(5)}$ | Binary nickel borides |

The magnetic energy of the altermagnetic ground state relative to the nonmagnetic state (or second-lowest energy FM state) increased from -62.1 (-2.8) meV/atom in the ordered phase (Figure 3a) to -42.4 (-2.6) meV/atom in the disordered model (Figure S10b). Therefore, although the mixed occupancy and partial site disorder can slightly decrease magnetic stability, our predicted altermagnetic ground state for $FeMo_2B_2$ is unchanged. Such a conclusion should also hold true for the altermagnetic $FeTa_2B_2$, which was observed to have a similar mixed occupancy situation, a fully occupied Ta site, and 6% vacancy on the Fe site (Table 4). The predicted magnetic ground state is not expected to change due to such a percentage of vacancy or substitution by nonmagnetic atoms. More importantly, the altermagnetic ground state in $FeTa_2B_2$ is more stable relative to the second-lowest energy FM state, -6.2 meV/atom (Figure 3a), compared to that for FeMo$_2$B$_2$. For FeNb$_2$B$_2$, the altermagnetic ground state is also very stable relative to the second-lowest energy FM state, -6.7 meV/atom (Figure 3a). However, additional Fe was observed to occupy the Nb site by 10% (Table 4). Although general altermagnetism is still expected in FeNb$_2$B$_2$, how much magnetism deviates from it requires a follow-up experimental investigation. For MnMo$_2$B$_2$, the altermagnetic ground state only differs from the second-lowest energy AAF state by -1.0 meV/atom (Figure 3a). 19% Mo occupancy on the Mn site (Table 4) may make MnMo$_2$B$_2$ suffer from magnetic fluctuations.

A previous experimental study[43] reported that FeMo$_2$B$_2$ exhibited FM characteristics with a saturation magnetization of 8.35 emu/g at room temperature[43]. This claim contradicts the theoretical ground state of FeMo$_2$B$_2$ in this study, reported $^{57}$Fe Mössbauer spectroscopy study showing no magnetic ordering for FeMo$_2$B$_2$ at room temperature[92], and another theoretical prediction[42]. Moreover, 8.35 emu/g is too small to claim ferromagnetism for typical Fe-containing FM materials, e.g., bcc Fe (~220 emu/g) and Fe$_2$B (156.9 emu/g[93]). Thus, the ferromagnetism may come from some FM impurities such as Fe or binary Fe borides.

A comprehensive experimental investigation into the magnetic properties of these phases is currently in progress. Determining the exact magnetic structure of these compounds requires neutron diffraction studies, which impose limitations on the boron source. Due to considerable neutron absorption of the $^{10}$B component of natural abundance boron, the neutron powder diffraction studies cannot utilize samples made with amorphous boron. Instead, samples must be re-synthesized using isotopically enriched crystalline $^{11}$B, which has a low neutron absorption. The optimized reaction conditions for amorphous boron differ from those needed for the less reactive crystalline $^{11}$B starting material. We are currently modifying the synthesis processes to produce single-phase samples of $^{11}$B-containing $TM_2$B$_2$, which will be suitable for neutron powder diffraction experiments. Synthesis of such samples and exploration of their magnetism is currently underway. The results of the complete magnetic characterization of these phases will be reported in due course.

**CONCLUSIONS**

In summary, utilizing first-principles calculations, we systematically studied the thermodynamic, electronic, and magnetic properties of $TM_2$B$_2$ ternary borides, where $T$ is 3$d$ and $M$ is 4$d$/5$d$ transition metals. We found 60 and 60 thermodynamically stable/metastable $TM_2$B$_2$ phases of the tetragonal FeMo$_2$B$_2$-type and the orthorhombic CoW$_2$B$_2$-type, respectively, within 0.2 eV/atom above the convex hull. Among them, we identified 24 and 16 magnetic phases in the tetragonal and orthorhombic structures, respectively. The rest are nonmagnetic.

Among the magnetic phases in the tetragonal structure, we discovered the first family of altermagnets in borides, which comprises 11 compositions: FeMo$_2$B$_2$, FeNb$_2$B$_2$, FeTa$_2$B$_2$, MnMo$_2$B$_2$, MnRu$_2$B$_2$, FeZr$_2$B$_2$, FeRu$_2$B$_2$, MnW$_2$B$_2$, MnRe$_2$B$_2$, FeHf$_2$B$_2$, and FeW$_2$B$_2$. The rest of the magnets in the tetragonal structure and all the magnets in the orthorhombic structure are conventional ferromagnets or antiferromagnets. The altermagnetism in FeNb$_2$B$_2$ and FeTa$_2$B$_2$ confirmed other recent reports for their band splitting. To the best of our knowledge, reports on the remaining 9 altermagnets, FeMo$_2$B$_2$, MnMo$_2$B$_2$, MnRu$_2$B$_2$, FeZr$_2$B$_2$, FeRu$_2$B$_2$, MnW$_2$B$_2$, MnRe$_2$B$_2$, FeHf$_2$B$_2$, and FeW$_2$B$_2$, are unavailable. The electronic structures of this family of altermagnets feature momentum-dependent band splitting for opposite spins. Their magnonic structures also feature momentum-dependent band splitting for opposite magnonic chiralities. The fundamental cause of both the altermagnetic splitting and chiral magnon splitting can be traced back to the unique arrangement of their magnetic structures, in which the nonmagnetic atoms play a pivotal role. The arrangement of the tetragonal *M*-prism coordinated to each *T* atom precludes the translation and inversion symmetries but allows only rotational or mirror symmetries to connect opposite-spin sublattices. This universally shared structural feature by the tetragonal phases gives birth to this family of altermagnets, which exist in considerable numbers. In contrast, with a different arrangement of the same set of atoms, the orthorhombic structure forms no altermagnet.

For these altermagnets, spin-splitter gauge factors are of order unity. Large anomalous Hall conductivity is predicted in the easy-plane FeNb$_2$B$_2$, FeTa$_2$B$_2$, and MnMo$_2$B$_2$. Large anomalous Hall conductivity is also expected in the spin-flop state for FeMo$_2$B$_2$ and FeW$_2$B$_2$.

The unique crystal feature required by altermagnetism and the existence of a competing phase pose a challenge for the synthesis of these altermagnets. To verify the theoretical predictions, we developed an iodine-assisted synthesis method for *TM*$_2$B$_2$ borides. Using the developed new method, we synthesized 4 tetragonal altermagnetic phases, FeMo$_2$B$_2$, FeNb$_2$B$_2$, FeTa$_2$B$_2$, MnMo$_2$B$_2$, and 3 orthorhombic nonmagnetic phases, CoMo$_2$B$_2$, NiMo$_2$B$_2$, and FeW$_2$B$_2$. To the best of our knowledge, 3 of the altermagnets, FeNb$_2$B$_2$, FeTa$_2$B$_2$, and MnMo$_2$B$_2$, were synthesized for the first time since they were briefly mentioned to exist more than half a century ago. For FeMo$_2$B$_2$, although recent syntheses exist, detailed information on the admixtures, especially magnetic impurities, was unavailable. Here, admixtures in the synthesized samples were carefully analyzed and documented, which facilitates future experimental magnetic characterizations. A feedback loop between theory and experiments was employed to improve the prediction models for partial site disorders, the implications of which have been discussed. Experimental magnetic property investigations combined with the synthesis of $^{11}$B isotopically enriched samples for neutron powder diffraction studies are currently being conducted and will be reported in due course.

This work enables the study of altermagnetism and altermagnons in borides. It also offers valuable insights into the discovery and design of altermagnets. We demonstrate that once a structural motif meets the symmetry requirements for altermagnetism, altermagnets can emerge as a family crystallizing in the same structure type. Consequently, we anticipate altermagnets to be much more abundant in nature than currently known. This study demonstrates the feasibility of both high-throughput computational screening and elemental substitutions in experiments to discover altermagnets.

ACKNOWLEDGMENTS

We are thankful to Dr. Jianming Bai and Dr. Hui Zhong at 28-ID-2 NSLS-II BNL for assisting with the *in-situ* PXRD data collection experiments. We also thank the staff scientist Dr. Kevin Stone at the BL2-1 of SLAC for his support with our mail-in proposal and the high-resolution PXRD data collection. This work was supported by the U.S. Department of Energy (DOE) Established Program to Stimulate Competitive Research (EPSCoR) Grant No. DE-SC0024284. The Ames National Laboratory is operated for the U.S. DOE by Iowa State University under Contract No. DE-AC02-07CH11358. Computations were performed at the High Performance Computing facility at Iowa State University and the Holland Computing Center at the University of Nebraska. Use of the Stanford Synchrotron Radiation Light Source, SLAC National Accelerator Laboratory, is supported by the U.S. DOE, Office of Science, Office of Basic Energy Sciences under Contract No. DE-AC02-76SF00515. This research used beamline 28-ID-2 of the National Synchrotron Light Source II, a U.S. DOE Office of Science User Facility operated for the DOE Office of Science by Brookhaven National Laboratory under Contract No. DE-SC0012704. Y.S. acknowledges the support by the National Natural Science Foundation of China (T2422016) and the Natural Science Foundation of Xiamen Municipality (3502Z202371007).

# Supporting Information

## Discovery and Synthesis of a Family of Boride Altermagnets

Zhen Zhang[1,#], Eranga H. Gamage[2,3,#], Genevieve Amobi[2], Subhadip Pradhan[4], Andrey Kutepov[4], Kirill D. Belashchenko[4], Yang Sun[5], Kirill Kovnir[2,3,*], and Vladimir Antropov[1,3,*]

[1] Department of Physics and Astronomy, Iowa State University, Ames, IA 50011, USA
[2] Department of Chemistry, Iowa State University, Ames, IA 50011, USA
[3] Ames National Laboratory, U.S. Department of Energy, Ames, IA 50011, USA
[4] Department of Physics and Astronomy and Nebraska Center for Materials and Nanoscience, University of Nebraska-Lincoln, Lincoln, NE 68588, USA
[5] Department of Physics, Xiamen University, Xiamen 361005, China
[#] These authors contributed equally.
[*] Corresponding authors: Kirill Kovnir kovnir@iastate.edu, Vladimir Antropov antropov@iastate.edu

## Contents

## Text S1. Computational Methods

**DFT Calculations.** We conducted DFT calculations using the projector-augmented wave method (PAW) as implemented in the VASP package[1]. The exchange-correlation energy was treated by the Perdew-Burke-Ernzerhof (PBE) GGA. A plane-wave basis set with a kinetic energy cutoff of 600 eV and a Gaussian smearing of 0.05 eV were used. The convergence thresholds were $10^{-5}$ eV for electronic self-consistency and 0.01 eV Å$^{-1}$ for ionic relaxation. A Γ-centered $2\pi \times 0.02$ Å$^{-1}$ k-point grid was used for the BZ sampling in the structural optimization, total energy, magnetic moment, and electronic density of states calculations.

**Magnetic Anisotropy Calculations.** The magnetic anisotropy calculations were performed using VASP with the PBE GGA functional. The magnetocrystalline anisotropy energy (MAE) was determined by evaluating the energy differences associated with spin orientations along in-plane and out-of-plane crystallographic directions. Initially, a collinear ground-state calculation was conducted to obtain the self-consistent charge density. This charge density was then used as input for non-self-consistent calculations that included spin-orbit coupling to compute the total energy for different spin orientations. A plane-wave energy cutoff of 600 eV was employed, and a Monkhorst k-point mesh of $22 \times 22 \times 40$ was used to ensure convergence of the MAE to within 0.1 meV per magnetic atom.

**Exchange Coupling Calculations.** The calculations of the exchange coupling parameters have been performed using a full-potential linearized augmented plane wave method with the FlapwMBPT[2,3] code. Our test results for ferromagnetic bcc Fe revealed a close agreement with the earlier publications. The implementation follows the original LCAO definition of the RKKY Heisenberg model parameters[4], we use the following integrals taken over the corresponding muffin-tin spheres,

$$J_{tt'}^{RR'} = -2 \int_{\Omega_{Rt}} d\mathbf{r} \int_{\Omega_{R't'}} d\mathbf{r}' B^{xc}(\mathbf{r})[\chi^{+-}(\mathbf{r},\mathbf{r}')]B^{xc}(\mathbf{r}'), \quad (S1)$$

where $B^{xc}(\mathbf{r})$ is the exchange-correlation spin-polarized field (the spin splitting in the original paper[4]), $\mathbf{R}, \mathbf{R}'$ are the translation vectors, and $\mathbf{t}, \mathbf{t}'$ represent the positions of atoms in the unit cell. $[\chi^{+-}(\mathbf{r},\mathbf{r}')]$ is the transversal static spin susceptibility matrix, which for collinear systems is presented as

$$\chi^{+-}(\mathbf{r},\mathbf{r}') = \sum_{mn} \frac{f_{n\uparrow} - f_{m\downarrow}}{\epsilon_{n\uparrow} - \epsilon_{m\downarrow}} \psi_{n\uparrow}^*(\mathbf{r})\psi_{m\downarrow}(\mathbf{r})\psi_{m\downarrow}^*(\mathbf{r}')\psi_{n\uparrow}(\mathbf{r}'), \quad (S2)$$

where $\psi_m, \psi_n$ are non-relativistic wave functions for the different spins. For the periodic systems the corresponding Fourier transform (with corresponding band indexes $\lambda, \lambda'$) is

$$\begin{aligned}
J_{tt'}^{\mathbf{q}} = &-\frac{2}{N_\mathbf{k}} \sum_\mathbf{k} \sum_{\lambda\lambda'} \\
&\times \int_{\Omega_t} d\mathbf{r} B^{xc}(\mathbf{r})\psi_\lambda^{\downarrow\mathbf{k}}(\mathbf{r})\psi_{\lambda'}^{*\uparrow\mathbf{k}-\mathbf{q}}(\mathbf{r}) \\
&\times \frac{f_{\lambda'}^{\uparrow\mathbf{k}-\mathbf{q}} - f_\lambda^{\downarrow\mathbf{k}}}{\epsilon_{\lambda'}^{\uparrow\mathbf{k}-\mathbf{q}} - \epsilon_\lambda^{\downarrow\mathbf{k}}} \\
&\times \int_{\Omega_{t'}} d\mathbf{r}' B^{xc}(\mathbf{r}')\psi_\lambda^{*\downarrow\mathbf{k}}(\mathbf{r}')\psi_{\lambda'}^{\uparrow\mathbf{k}-\mathbf{q}}(\mathbf{r}').
\end{aligned} \quad (S3)$$

In order to evaluate the magnon energies (see below) and plot them along some directions in the k-space, we have to be able to interpolate the exchange parameters from our fixed k-mesh onto an arbitrary point in the reciprocal space. This is done using Fourier interpolation. First, we apply the Fast Fourier Transform to get the exchange parameters in the active space of the **R**-vectors,

$$J_{tt'}^{\mathbf{R}} = \frac{1}{N_{\mathbf{k}}} \sum_{\mathbf{q}} e^{i\mathbf{qR}} J_{tt'}^{\mathbf{q}}. \tag{S4}$$

$N_{\mathbf{k}}$ is the total number of k-points that we use to sample the Brillouin zone. Then we select only those **R**-vectors that are inside the sphere inscribed in the active polygon. We use these selected **R**-vectors to perform the Fourier transform back to the reciprocal space and get the exchange parameters for arbitrary **q**-points along our selected directions in the reciprocal space,

$$J_{tt'}^{\mathbf{q}} = \sum_{\mathbf{R}} e^{-i\mathbf{qR}} J_{tt'}^{\mathbf{R}}. \tag{S5}$$

For the spin wave spectrum of the systems with many magnetic atoms, we use the classical Heisenberg model result and diagonalize the following matrix,

$$\frac{2}{|\mathbf{M}_t|} \left\{ \frac{\mathbf{M}_t J_{tt'}^{\mathbf{q}}}{|\mathbf{M}_{t'}|} - \delta_{tt'} \sum_{t''} J_{tt''}^{\mathbf{q}=0} \frac{\mathbf{M}_{t''}}{|\mathbf{M}_{t''}|} \right\}. \tag{S6}$$

**Text S2. Experimental Methods**

**Chemical Materials:** Boron powder (Amorphous & crystalline, -325 mesh, 98%, Thermo Scientific), Nb powder (99.8%, -325 mesh, Thermo Scientific), Mo powder (99.5%, Alfa Aesar), Ta powder (99.97%, -325 mesh, Thermo Scientific), W powder (99.9%, 1-5 micron, Alfa Aesar), Fe powder (0.4 mmol, <10 μm, Alfa Aesar, 99.5%), Co powder (99.8%, Alfa Aesar), Mn powder (99.95%, -325 mesh, Alfa Aesar), Ni powder (99.9%, -325 mesh, Thermo Scientific), and $I_2$ crystals (99.999%, Sigma Aldrich) were used as received from the vendors without any pre-treatment.

**Synthesis:** Bulk powders of different $TM_2B_2$ ($T$ = Fe, Co, Mn, Ni; $M$ = Nb, Mo, Ta, W) phases were synthesized in an iodine-mediated solid-state synthesis reaction. Samples were prepared by loading the required amounts of transition metals and boron powders into a silica ampoule. The ampoules were covered with parafilm and placed in an ice bath before adding $I_2$ crystals to minimize the evaporation of iodine upon evacuation and sealing. After sealing, the ampoules were wrapped in silica wool and placed in a muffle furnace at room temperature. The furnace was ramped up to 950-1100 °C over 5-10 hours, dwelt at the set temperature for 24-72 hours, and cooled to room temperature over 5 hours. The ampoules were then opened in a fumehood, letting the iodine vapors escape. The contents were transferred to glass beakers that were subsequently filled with ~50 mL of dilute HCl (pH~4) solution and kept under acid for 3-5 hours until gas bubbles no longer evolved. This washing removes recrystallized iodine, leftover $3d$ transition metals, and metal iodides. Overnight treatment in acid did not result in significant changes. Finally, the samples were filtered and dried under ambient conditions overnight before analysis. The target $TM_2B_2$ borides are air- and moisture-stable and were stored in vials in a desiccator in the long run.

**Characterization:** Powder X-ray diffraction (PXRD) was performed using a Rigaku Miniflex 600 with Cu-$K_\alpha$ radiation ($\lambda$ = 1.5406 Å) and a Ni-$K_\beta$ filter (Figures 7 and 8 of the main text). High-resolution powder X-ray diffraction experiments (HR-PXRD) at room temperature were conducted at the beamline BL2-1 of SLAC National Accelerator Laboratory, Stanford Synchrotron Radiation Light Source (SSRL). The energy of the X-rays employed was 17 keV ($\lambda$ = 0.72964 Å). The samples were diluted with silica powder to minimize X-ray absorption ($\mu R$ ~1-2) and packed in Kapton capillaries of ~0.7 mm outer diameter and <2 mm height. $LaB_6$ was used as the calibration standard. Rietveld refinement was performed on PXRD data using the GSAS-II software package (version 5804)[5]. The background was first refined using the fixed-point method, followed by the unit cell refinement, sample displacement, size, microstrain, and preferred orientation. Next, the atomic coordinates and atomic displacement parameters (ADPs) were refined for the transition metals. Boron coordinates were not refined, and the ADPs were fixed to be close to those refined for transition metals. At the final stage of the refinement, the mixed occupancy in metal atomic positions was probed under constraints of equivalent atomic coordinates and ADPs, and the total occupancy of each site was equal to 100%. For $FeTa_2B_2$, the electron density in the Fe site was statistically lower than 100% of Fe. The resulting $R_w$ values of the refinements with and without mixed occupancy were compared.

Variable-temperature *in-situ* X-ray diffraction experiments were conducted with 0.1811 Å wavelength synchrotron radiation at the beamline 28-ID-2 of the National Synchrotron Light Source II (NSLS-II) at Brookhaven National Laboratory. The experimental setup featured a horizontal flow cell, with the capillary affixed between two heating coils on either side. Boron and transition metals were ground into a homogeneous mixture, packed in silica capillaries (1.1 mm OD/0.9 mm ID) sandwiched between iodine crystals (Figure S6), and flame-sealed under vacuum. The heating and cooling of the sample were conducted according to a preset reaction profile. The X-ray beam was directed at the powder in the middle, and an IR sensor was used to read the temperature of the sample during the experiment. 2D diffraction images were collected every 60 s and were converted to 1D diffraction patterns for analysis through integration using GSAS-II.

Elemental analysis of samples was conducted by energy dispersive X-ray spectroscopy (EDX) coupled with scanning electron microscopy (SEM) technique on an FEI Quanta 250 field emission-SEM with an EDX detector (Oxford X-Max 80, ThermoFischer Scientific, Inc., USA). The experiments were performed, and data were analyzed using Aztec software. Powder samples were sprinkled on carbon tape to obtain quants. Accelerating voltages of 5 kV and 15 kV were used to study all samples. Accurate quantitative analysis of boron cannot be performed by EDX as the light boron element does not provide reliable quants. However, the composition of the heavy $3d$/$4d$/$5d$ elements could be determined successfully as their X-ray emission lines differ from those of boron. The normalized EDX data obtained for the $TM_2B_2$ phases is in good agreement with the refined compositions (Table 4 of the main text). Moreover, EDX/SEM was useful in determining the different types of admixtures in the samples.

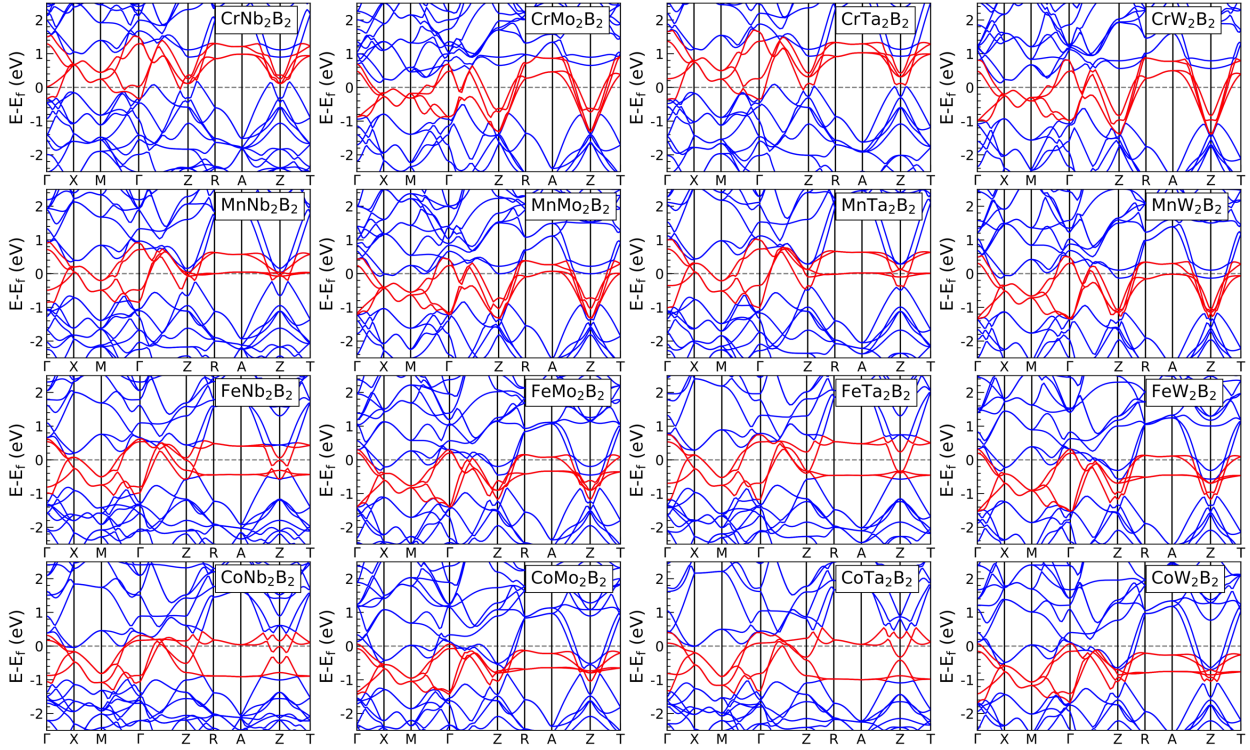

**Figure S1.** Nonmagnetic band structure of the tetragonal $TM_2B_2$ ($T$ = Cr, Mn, Fe, Co; $M$ = Nb, Mo, Ta, W). High-symmetry points in the Brillouin zone are denoted as follows: $\Gamma(0, 0, 0)$, $X(0, 1/2, 0)$, $M(1/2, 1/2, 0)$, $Z(0, 0, 1/2)$, $R(0, 1/2, 1/2)$, $A(1/2, 1/2, 1/2)$, $T(1/2, 1/4, 1/2)$. Red color highlights nearly flat bands along Z-R-A-Z-T corresponding to two narrow peaks in the nonmagnetic DOS.

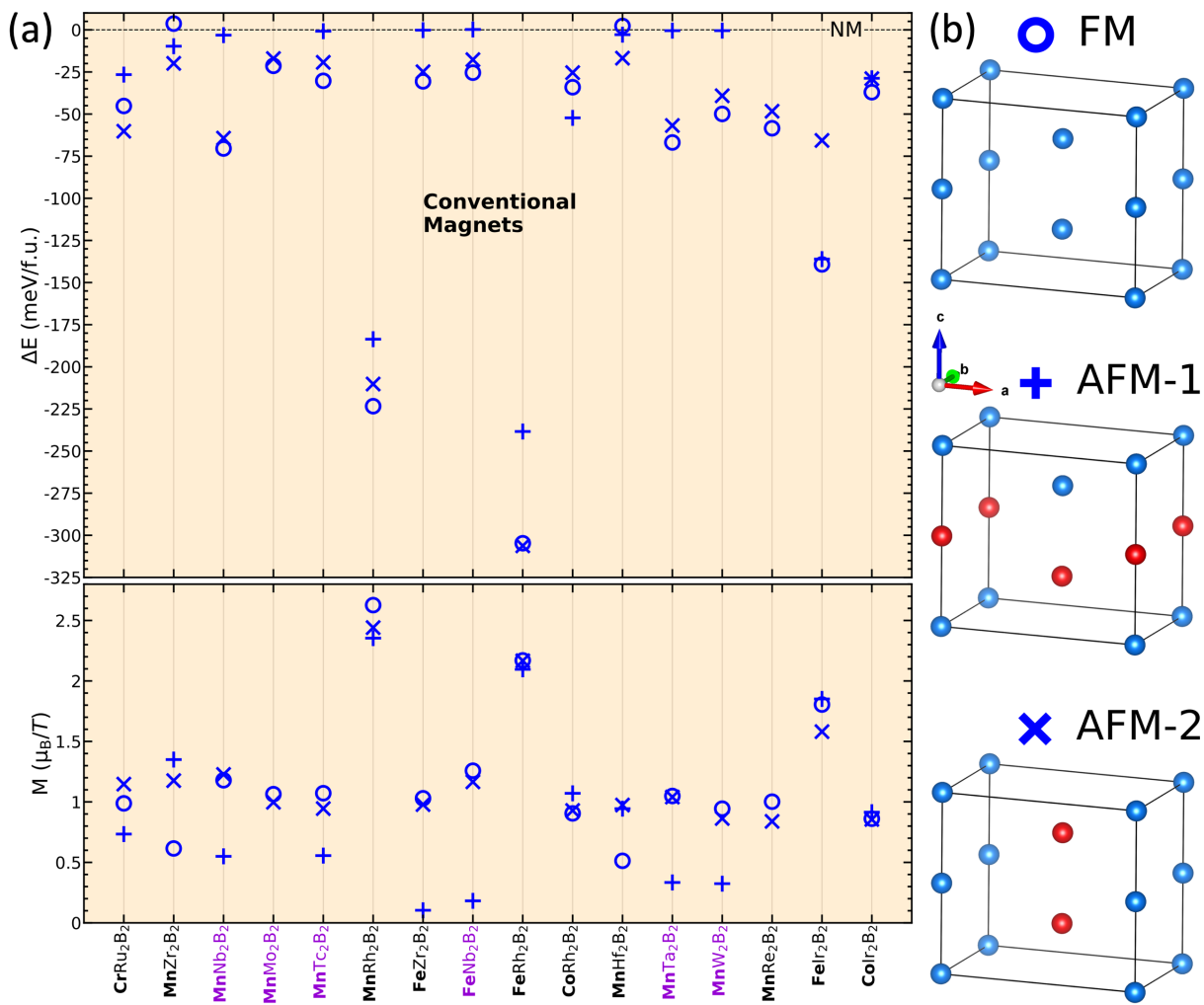

**Figure S2.** Magnetic solutions for the orthorhombic $TM_2B_2$ phases within 0.2 eV/atom above the convex hull. (a) The relative energy difference of the magnetic solutions with respect to the nonmagnetic one (top panel) and the magnetic moment on the $T$ atom (bottom panel). Purple compositional x-labels indicate compounds within 0.05 eV/atom above the convex hull. Black compositional x-labels indicate compounds within 0.2 eV/atom above the convex hull. (b) Different magnetic configurations within the 20-atom $1 \times 1 \times 2$ supercell and the associated symbols and labels. Blue and red spheres indicate $T$ atoms with opposite spins. $M$ and B atoms are not shown.

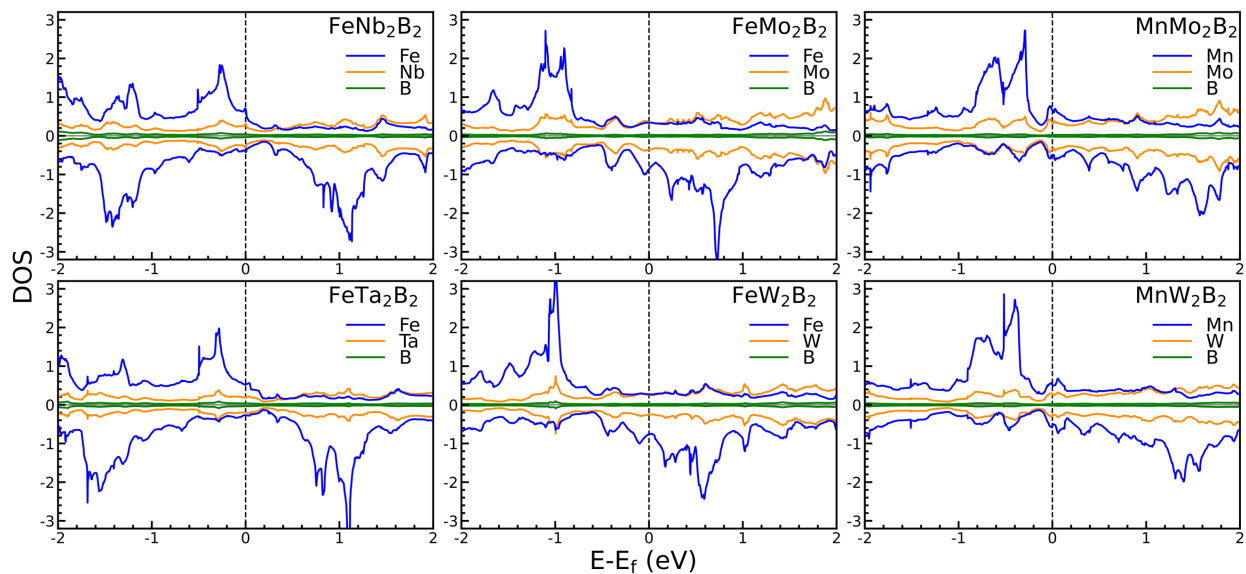

**Figure S3.** Spin-polarized electronic DOS (eV$^{-1}$ atom$^{-1}$ spin$^{-1}$) of the magnetic ground state of the tetragonal altermagnets, FeNb$_2$B$_2$, FeTa$_2$B$_2$, FeMo$_2$B$_2$, FeW$_2$B$_2$, MnMo$_2$B$_2$, and MnW$_2$B$_2$. Blue, orange, and green curves represent partial DOS (PDOS) of a single *T*, *M*, and B atom, respectively. Up and down spins represent the majority and minority spins, respectively, for an atom.

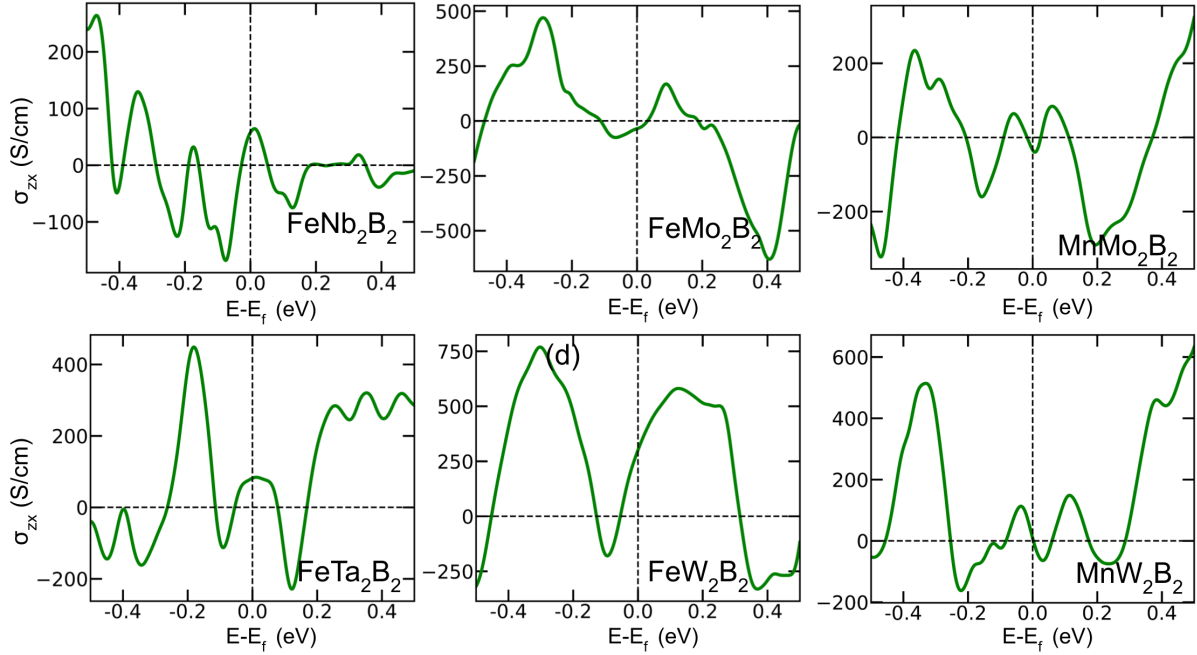

**Figure S4.** Anomalous Hall conductivity (AHC) component $\sigma_A^{zx}$ as a function of the Fermi energy for the Néel vector oriented along the [100] crystallographic direction for the tetragonal altermagnets.

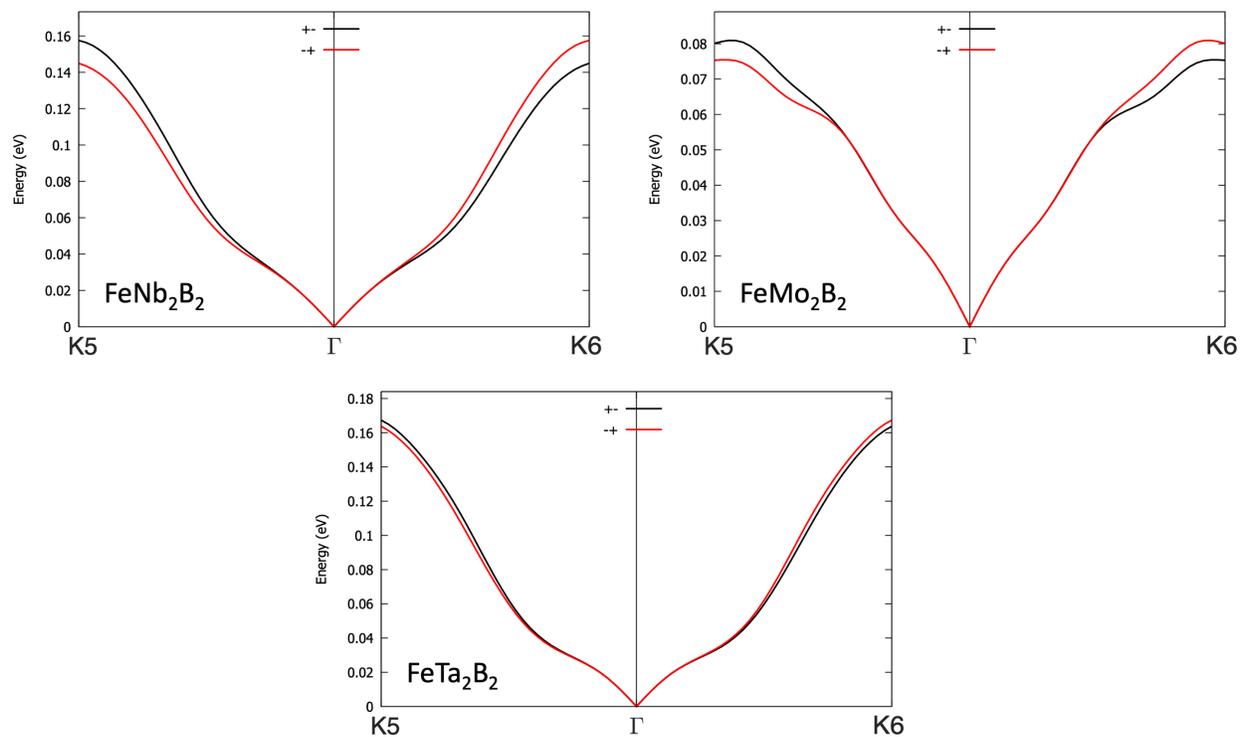

**Figure S5.** Spin-wave spectra out of the $k_z = 0$ plane of the tetragonal FeNb$_2$B$_2$, FeMo$_2$B$_2$, and FeTa$_2$B$_2$ by GGA. Red and black curves indicate opposite magnon chiralities. Non-high-symmetry points are denoted as follows: K5(1/3, 1/6, 1/2), K6(-1/3, 1/6, 1/2).

**Table S1.** Optimized reaction conditions to form $TM_2B_2$ phases. The end products were determined by *ex-situ* PXRD and EDX. Some trace admixtures, which were not detected in PXRD, were identified using EDX. The reactions were conducted on a 100-200 mg scale. *Optimized ratios that produced ≥90% of the target ternary boride. #All the reactions were cooled down to 20 °C over 5 hours.

| Target Phase | Loading ratios* | Ramp time/ dwell temp/ dwell time# | Minor admixtures identified by PXRD+EDX |
|---|---|---|---|
| $FeMo_2B_2$ | $2Mo+2Fe+2B+1.5I_2$ | 10h/1050°C/48h | FeB, $Fe_2B$ |
| $MnMo_2B_2$ | $1.5Mo+4.5Mn+2B+2I_2$ | 10h/950°C/48h | Cristobalite (α-$SiO_2$), MoB |
| $FeNb_2B_2$ | $2Nb+2.5Fe+2B+1.5I_2$ | 5h/1100°C/24h | Binary niobium borides |
| $FeTa_2B_2$ | $2Ta+1.75Fe+2B+1.5I_2$ | 5h/1100°C/24h | TaB, $Ta_2O_5$ |
| $CoMo_2B_2$ | $2Mo+2.5Co+2B+3I_2$ | 10h/950°C/48h | CoB, $Co_2B$ |
| $NiMo_2B_2$ | $2Mo+1.5Ni+2B+1.5I_2$ | 5h/1100°C/24h | Binary nickel borides |
| $FeW_2B_2$ | $1.5W+3.5Fe+2B+2I_2$ | 10h/1050°C/48h | FeB, $Fe_2B$ |

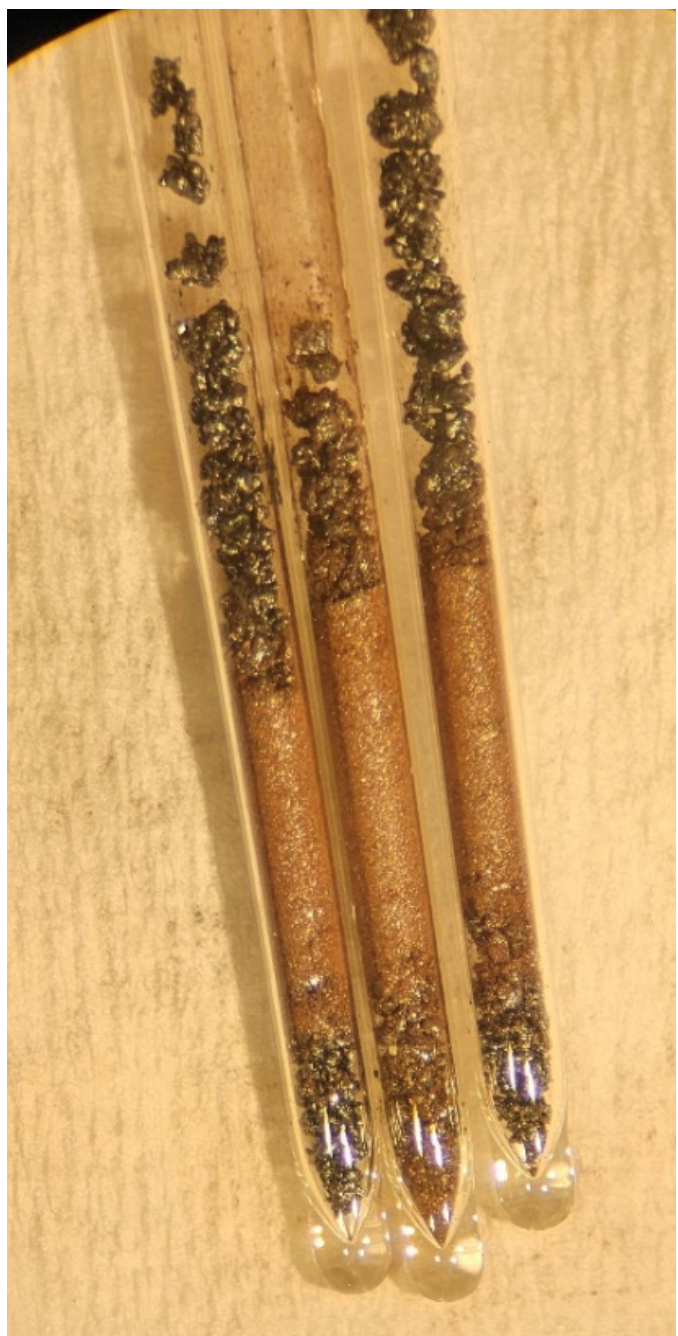

**Figure S6.** Silica reaction capillaries (0.9 mm ID/1.1 mm OD) packed with finely ground Fe+Mo+B powders sandwiched between iodine crystals prepared for variable temperature *in-situ* PXRD experiments at beamline 28-ID-2 of the National Synchrotron Light Source II (NSLS-II) at Brookhaven National Laboratory.

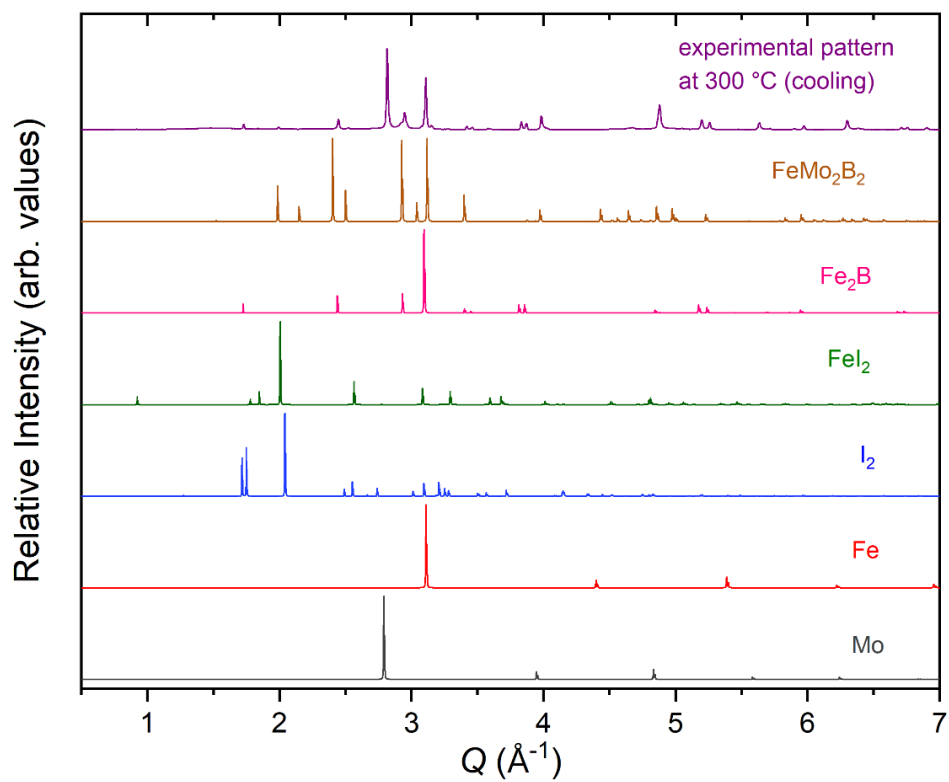

**Figure S7.** Reference PXRD patterns of the associated phases shown in the 3D waterfall plot of the *in-situ* solid-state reaction conducted at the beamline 28-ID of the NSLS-II at BNL in Figure 9 of the main text. Please note that the shift in peak positions in the experimental pattern is due to thermal expansion effects.

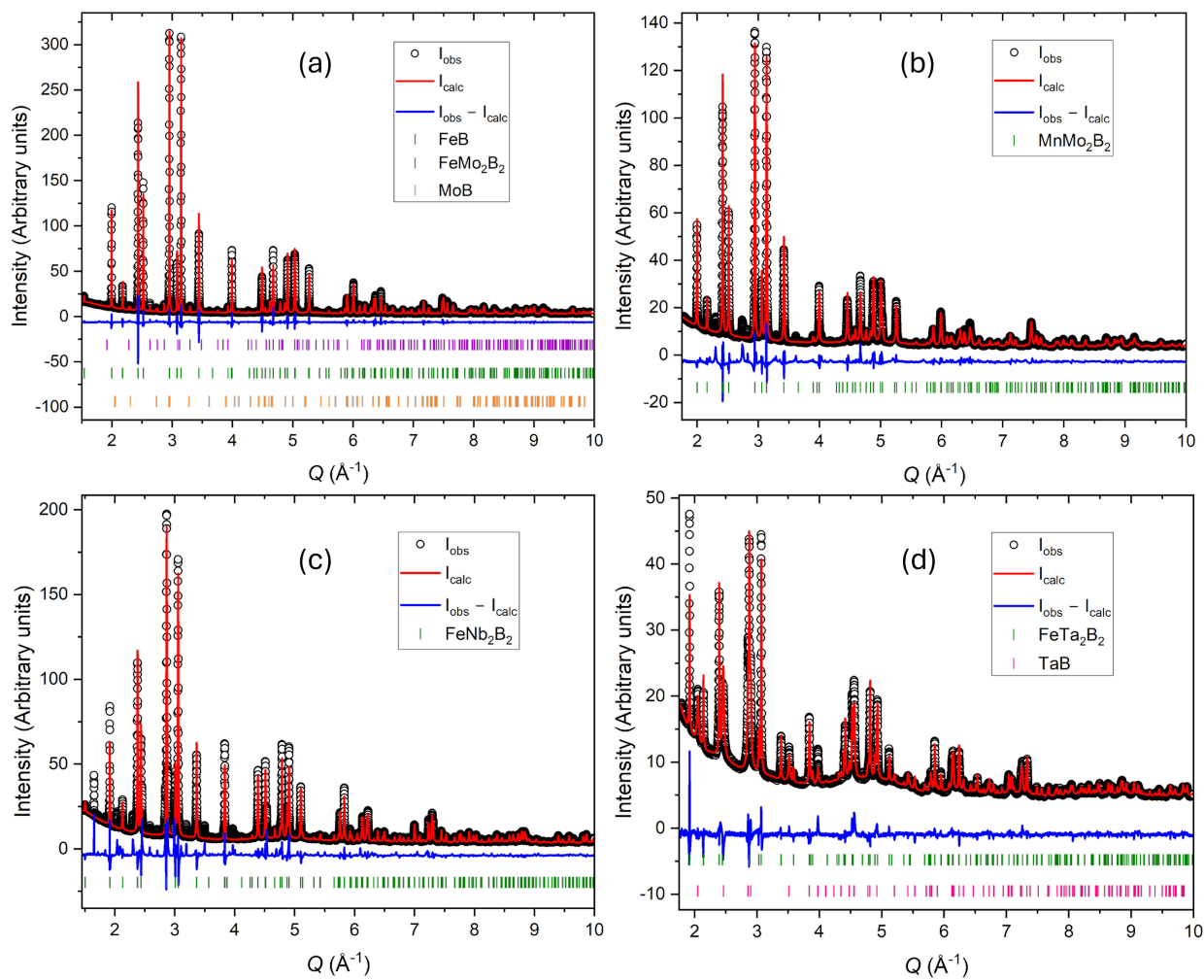

**Figure S8.** Rietveld refinement performed on the synchrotron HR-PXRD ($\lambda = 0.72964$ Å) data of the tetragonal (a) FeMo$_2$B$_2$, (b) MnMo$_2$B$_2$, (c) FeNb$_2$B$_2$, and (d) FeTa$_2$B$_2$ samples collected at 300 K. In (b) and (c), the proper composition of the binary/ternary admixture was not identified. Hence, admixtures were not included in the refinement. Note that the samples sent for synchrotron PXRD measurements were not fully optimized, and the refinements should not be used to assess phase purity.

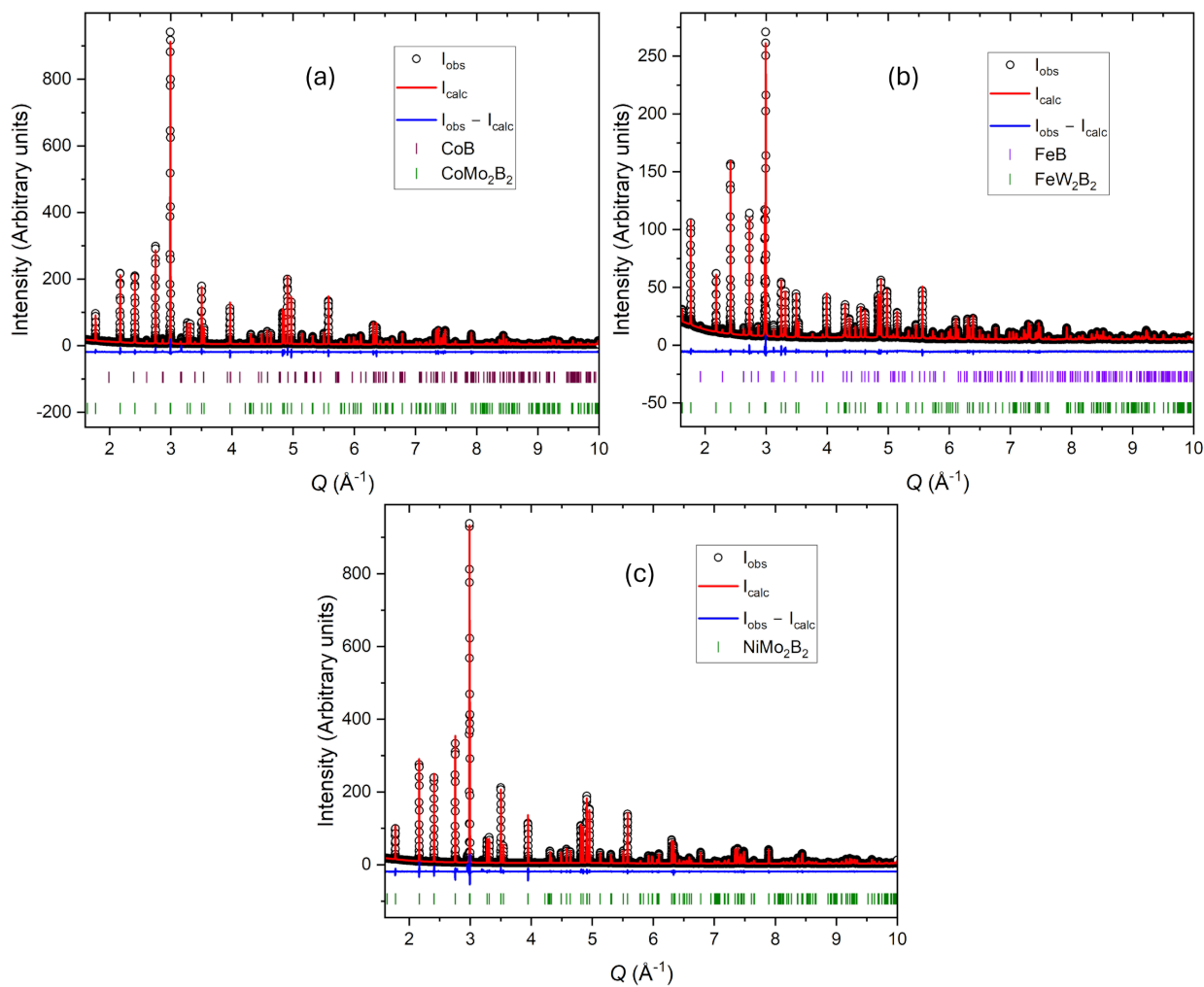

**Figure S9.** Rietveld refinement performed on the synchrotron HR-PXRD ($\lambda$ = 0.72964 Å) data of the orthorhombic (a) $CoMo_2B_2$, (b) $FeW_2B_2$, and (c) $NiMo_2B_2$ samples collected at 300 K. In (c), the proper composition of the binary admixture was not identified. Hence, admixtures were not included in the refinement. Note that the samples sent for synchrotron PXRD measurements were not fully optimized, and the refinements should not be used to assess phase purity.

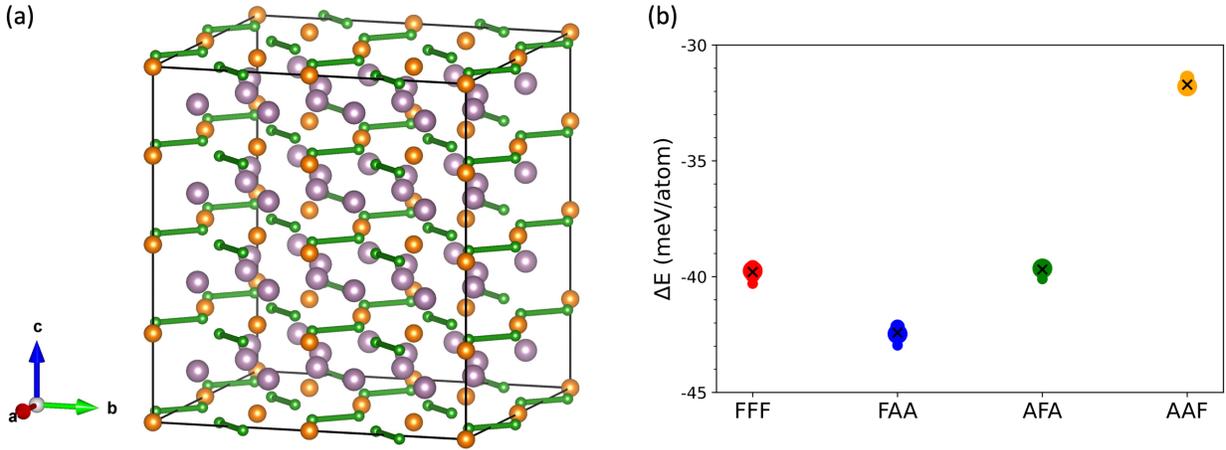

**Figure S10.** (a) 2 × 2 × 4 supercell (160 atoms) of the tetragonal FeMo$_2$B$_2$ structure containing 32 Fe atoms used for the disordered model. Fe: orange, Mo: purple, B: green. (b) The relative energy difference of the magnetic solutions with respect to the nonmagnetic one. The scatters of each magnetic configuration indicate the variation of energies for inequivalent disordered supercell configurations. The size of each scatter indicates the relative weight of each inequivalent supercell. The cross mark indicates the weight-averaged energy over all inequivalent supercells, which is the energy of each magnetic configuration obtained by the constructed disordered model.